\documentclass[a4paper,11pt,reqno]{article}

\usepackage{a4wide}
\usepackage{amsmath,amsfonts,amssymb}
\usepackage[english]{babel}
\usepackage{soul}
\usepackage{nicefrac}
\usepackage[mathscr]{euscript}
\usepackage{setspace}
\usepackage{datetime}
\usepackage[sort&compress,merge,numbers]{natbib}
\usepackage{mathrsfs}
\usepackage[T1]{fontenc}
\usepackage{mathpazo}

\usepackage[breaklinks=true]{hyperref}

\numberwithin{equation}{section}

\hypersetup{
			colorlinks=true,
			urlcolor=blue,
			citecolor=magenta,
			linkcolor=blue,
     }

\let\oldsqrt\sqrt
\def\sqrt{\mathpalette\DHLhksqrt}
\def\DHLhksqrt#1#2{%
\setbox0=\hbox{$#1\oldsqrt{#2\,}$}\dimen0=\ht0
\advance\dimen0-0.2\ht0
\setbox2=\hbox{\vrule height\ht0 depth -\dimen0}%
{\box0\lower0.4pt\box2}}

\author{
  \begin{minipage}{.97\linewidth}
    \vspace{1cm}
       \begin{center}
      \begin{small}
               \textbf{Luca Ciambelli},$^1$ 
             \textbf{Anastasios C. Petkou},$^2$ 
                     \\
     \textbf{P. Marios Petropoulos}$^1$ and 
      \textbf{Konstantinos Siampos}$^3$
              \end{small}
    \end{center}
    \vspace{0.5cm}
    \hspace{2.4cm}\begin{minipage}{.7\linewidth}
\begin{center}     {\it \begin{footnotesize}
\hbox{\kern-1.8cm\vbox{\vskip0cm
 \begin{itemize}
               \item[$^1$] CPHT -- Centre de Physique Th\'eorique\\ 
        Ecole Polytechnique, CNRS UMR 7644\\
        Universit\'e Paris--Saclay\\
        91128 Palaiseau Cedex, France
\vskip0.3cm
      \end{itemize}}
\kern-3cm\vbox{
\begin{itemize}
 \item[$^2$] Department of Physics\\ 
  Institute of Theoretical Physics\\
  Aristotle University of Thessaloniki\\ 
  54124, Thessaloniki, Greece
      \end{itemize}\vskip0.05cm
}}
     \end{footnotesize}}
\end{center}
    \end{minipage}
    \vspace{0.5cm}\begin{minipage}{.7\linewidth}
\begin{center}     
{\it \begin{footnotesize}
\hbox{\kern3.5cm\vbox{\vskip0cm
 \begin{itemize}
             \item[$^3$] Albert Einstein Center for Fundamental Physics\\
Institute for Theoretical Physics\\ 
University of Bern\\
Sidlerstrasse 5, 3012 Bern, Switzerland
\vskip0.29cm
      \end{itemize}}
}
     \end{footnotesize}}
\end{center}
     \end{minipage}
  \end{minipage}
}

\title{\vspace{2.5cm}
 \boldmath \begin{Large}
    \textbf{The Robinson--Trautman spacetime and its holographic fluid}
  \end{Large} \unboldmath
}

\date{}

\begin{document}

\begin{titlepage}
\maketitle
\thispagestyle{empty}

 \vspace{-14.cm}
  \begin{flushright}
  CPHT-PC037.062017\\
  \end{flushright}
 \vspace{12.5cm}

\begin{center}
\textsc{Abstract}\\  
\vspace{1cm}	
\begin{minipage}{1.0\linewidth}

We discuss the holographic reconstruction of four-dimensional asymptotically anti-de Sitter 
Robinson--Trautman spacetime from boundary data. We use for that a resummed version of the 
derivative expansion. The latter involves a vector field, which is interpreted as the dual-holographic-fluid velocity field and is naturally defined in the Eckart frame. In this 
frame the analysis of the non-perfect holographic energy--momentum tensor is considerably simplified.
The  Robinson--Trautman fluid is at rest and its time evolution is a heat-diffusion kind of phenomenon: 
the  Robinson--Trautman equation plays the r\^ole of heat equation, and the heat current is identified 
with the gradient of the extrinsic curvature of the two-dimensional boundary spatial section hosting 
the conformal fluid, interpreted as an  out-of-equilibrium kinematical temperature. The hydrodynamic-frame-independent entropy current is conserved for vanishing chemical potential, and the evolution of the fluid  resembles a Moutier thermodynamic path. We finally comment on the general transformation rules for moving to the Landau--Lifshitz frame, and on possible drawbacks of this option.

\end{minipage}
\end{center}


\end{titlepage}

\onehalfspace

\begingroup
\hypersetup{linkcolor=black}
\tableofcontents
\endgroup
\noindent\rule{\textwidth}{0.6pt}

\section{The Robinson--Trautman spacetime and holography}

Robinson--Trautman solutions to Einstein's equations were found in 1960-1962 \cite{RTorig}.\footnote{See e.g. \cite{GP} for a modern and more general presentation.}  They are obtained 
assuming the existence of a null, geodesic and shearless congruence. In vacuum, under these 
assumptions, Goldberg--Sachs theorem states that the corresponding spacetime is algebraically
 special, \emph{i.e.} Petrov type II, III, N, D or O. This feature remains valid when a
 cosmological constant or even certain other classes of energy sources are added. 

Asymptotically anti-de Sitter Robinson--Trautman spacetimes have attracted some attention in the 
framework of holography. The three-dimensional boundary metric and the dual conformal field theory expectation value of the energy--momentum tensor were found in \cite{deFreitas:2014lia}, 
where further properties of the boundary state were also discussed, in particular from a hydrodynamic 
perspective (see also \cite{Bakas:2014kfa}).

Conformal fluid dynamics was  thoroughly studied within fluid/gravity correspondence  
 \cite{Hubeny:2011hd, Haack:2008cp, Bhattacharyya:2008jc}.  This holographic correspondence 
sets a relationship between Einstein spaces (possibly with a gauge field) and boundary conformal 
fluids (potentially charged), incarnated in the derivative expansion. The derivative expansion 
is an alternative to the Fefferman--Graham expansion \cite{PMP-FG1, PMP-FG2}. Besides the usual
 boundary data as the metric and the energy--momentum tensor (for pure gravity), it requires an 
extra piece, namely a velocity field assumed to slowly vary in spacetime. 

In fact, the velocity field is redundant since it is not needed in the Fefferman--Graham approach, and it is 
arbitrary because for non-perfect relativistic fluids the distinction between energy and mass is immaterial.
 Its r\^ole is to organize the expansion, and its choice a matter of convenience, or better, of physical framework. 
 Often the derivative expansions are asymptotic series, and non-hydrodynamic (\emph{i.e.} non-perturbative) modes 
can appear, triggering an alarm regarding the validity of the hydrodynamic interpretation. From this viewpoint, 
some hydrodynamic-frame (velocity-field) choices might be better designated than others. 

Fluid/gravity correspondence raises an important question: given a boundary metric, what are the conditions it 
should satisfy, and which energy--momentum tensor should it be accompanied with in order for an \emph{exact} 
dual bulk Einstein space to exist? This question has been successfully investigated in 
\cite{Caldarelli:2012cm,Mukhopadhyay:2013gja, Petropoulos:2014yaa,Gath:2015nxa,Petropoulos:2015fba,Petkou:2015fvh}. 
It turns out to be relevant both for the integrability of Einstein's equations (\emph{\`a la Geroch}, see
 \cite{Geroch,Ehlers,Leigh:2014dja}) and because it gives access to exact transport properties of the 
holographic fluid. To answer this question the Fefferman--Graham expansion is not very useful because it 
is not resummable (except for trivial cases \cite{SS}), as opposed to the derivative expansion, which is 
resummable when the velocity field is chosen \emph{shearless}. 

The resummation process at hand reveals two main features: (\romannumeral1) the bulk Einstein spacetime
 is Petrov algebraically special, and (\romannumeral2) the boundary fluid velocity is in the Eckart frame. 
This last property is interesting because, often, the general analysis of transport properties in relativistic
 fluids is performed in the Landau--Lifshitz frame, hence setting the heat 
flow to zero. In the present framework, however, this choice is not natural, and can even be questionable. This happens in particular  for Robinson--Trautman spacetimes, which are algebraically special and emerge while resumming 
appropriate boundary data, and hence fall in the class under investigation here. In the following, we will 
review how Robinson--Trautman is obtained exclusively from boundary considerations (Sec. \ref{sec:RTres}), 
and what is the corresponding holographic-fluid interpretation, with some emphasis on the issue of entropy (Sec. \ref{sec:RTfluid}). Two appendices 
provide further useful information on relativistic hydrodynamics. 

\section{Reconstruction from the boundary}\label{sec:RTres}

Our aim here is to review the holographic construction of Robinson--Trautman Einstein spaces as
 performed in \cite{Gath:2015nxa}. We only refer to boundary data, which are designed and combined 
in order for the derivative expansion to be resummable. 

\subsection{The general resummation formula} \label{sec:resform}

If $\text{d}s^2=g_{\mu\nu}\text{d}x^\mu\text{d}x^\nu$ is the boundary metric and 
$\text{T}=T_{\mu \nu}\text{d}x^\mu\text{d}x^\nu$ is the boundary energy--momentum tensor, 
the resummed bulk metric\footnote{We have traded here the usual advanced-time coordinate used
 in the quoted literature on fluid/gravity correspondence for the retarded time, spelled $t$ (see \eqref{ut}).\label{rminusr}} 
reads:
\begin{equation}
\text{d}s^2_{\text{res.}} =
2\text{u}(\text{d}r+r \text{A})+r^2k^2\text{d}s^2+\frac{\Sigma}{k^2}
+ \frac{\text{u}^2}{\rho^2} \left(\frac{8\pi G T_{\lambda \mu}u^\lambda u^\mu}{k^2 }r+
\frac{C_{\lambda \mu}u^\lambda \eta^{\mu\nu\sigma}\omega_{\nu\sigma}}{2k^6}\right).
\label{papaefgenresc}
\end{equation}
\begin{itemize}
\item Here, $\text{u}$ is a shearless, normalized, time-like vector field. It has acceleration
 $\text{a}=\left(\text{u} \cdot \nabla\right)  \text{u}$,  expansion $\Theta = \nabla \cdot \text{u}$, and
vorticity $\omega=\frac{1}{2}\omega_{\mu\nu }\, \mathrm{d}\mathrm{x}^\mu\wedge\mathrm{d}\mathrm{x}^\nu  
=\frac{1}{2}\left(\mathrm{d}\mathrm{u} +
\mathrm{u} \wedge\mathrm{a} \right)$.

\item The guideline for setting up the derivative expansion is \emph{Weyl covariance} 
\cite{Haack:2008cp, Bhattacharyya:2008jc}: the bulk geometry is required to be insensitive to a 
conformal transformation of the boundary metric. Covariantization with respect to rescalings is 
achieved with the Weyl connection one-form:
\begin{equation}
\label{Wconc}
\text{A}=\text{a} -\frac{\Theta}{2} \text{u}.
\end{equation}
Covariant derivatives $\nabla$ are thus traded for Weyl-covariant ones $\mathscr{D}=\nabla+w\,\text{A}$, $w$ 
being the conformal weight of the tensor under consideration. In three spacetime dimensions, Weyl-covariant 
quantities are e.g. 
\begin{eqnarray}
\mathscr{D}_\nu\omega^{\nu}_{\hphantom{\nu}\mu}&=&\nabla_\nu\omega^{\nu}_{\hphantom{\nu}\mu},
\\
\mathscr{R}&=&R +4\nabla_\mu A^\mu- 2 A_\mu A^\mu , \label{curlRc}
\\
\mathscr{D}_\mu u_{\nu}&=&\nabla_{\mu}u_{\nu}+u_{\mu}a_{\nu}-\dfrac{\Theta}{2}h_{\mu\nu}\nonumber\\&=&\sigma_{\mu\nu}+\omega_{\mu\nu}
\end{eqnarray}
(for the last we have used \eqref{def1}), while
\begin{equation}
\label{sigmac}
\Sigma=
\Sigma_{\mu\nu} 
\text{d}x^\mu\text{d}x^\nu=-2\text{u}\mathscr{D}_\nu \omega^\nu_{\hphantom{\nu}\mu}\text{d}x^\mu- 
\omega_\mu^{\hphantom{\mu}\lambda} \omega^{\vphantom{\lambda}}_{\lambda\nu}\text{d}x^\mu\text{d}x^\nu
-\text{u}^2\frac{\mathscr{R}}{2} ,
\end{equation}
is Weyl-invariant and stands for the Weyl-covariantized Schouten tensor.

\item The radial coordinate is $r$, and $\rho$ performs the resummation of the derivative expansion as it is defined by
\begin{equation}\label{rho2c}
 \rho^2=r^2 +\frac{1}{2k^4} \omega_{\mu\nu} \omega^{\mu\nu} = r^2 +\frac{q^2}{4k^4}.
\end{equation} 
Boundary Weyl transformations $\text{d}s^2\to \nicefrac{\text{d}s^2}{{\cal B}^2}$
correspond to bulk diffeomorphisms, which can be reabsorbed into a redefinition of the radial coordinate:  $r\to{\cal B}\, r$.
 
\item The boundary metric $\text{d}s^2=g_{\mu\nu}\text{d}x^\mu\text{d}x^\nu$ has in general non-vanishing Cotton tensor 
$\text{C}=C_{\mu \nu}\text{d}x^\mu\text{d}x^\nu$, where
\begin{equation}
\label{cotdef}
C_{\mu\nu}=\eta_{\mu\rho\sigma}
\nabla^\rho \left(R_{\nu}^{\hphantom{\nu}\sigma}-\frac{R}{4}\delta_{\nu}^{\hphantom{\nu}\sigma} \right),
\end{equation}
with $\eta_{\mu\nu\sigma}=\sqrt{-g} \epsilon_{\mu\nu\sigma}$.
Whenever $\text{C}$ is non-zero, the bulk is asymptotically \emph{locally} anti-de Sitter. 
The Cotton tensor has conformal weight one (like the energy--momentum tensor) and is identically conserved:
\begin{equation}
\label{C-cons}
\nabla\cdot \text{C}=0.
    \end{equation}

\end{itemize}

The bulk metric $\text{d}s^2_{\text{res.}}$ given in expression \eqref{papaefgenresc} is an \emph{exact} Einstein 
space with $\Lambda=-3 k^2$
provided the boundary energy--momentum tensor is \emph{exactly} conserved: 
\begin{equation}
\label{T-cons}
\nabla\cdot \text{T}=0.
    \end{equation}
 This statement might raise questions, and calls for a few remarks.  The energy--momentum tensor is not meant 
to be necessarily of perfect-fluid type. At the same time, the time-like congruence $\text{u}$, chosen independently,
 is interpreted as the fluid velocity. It is somehow puzzling that despite the apparent (and, as we already 
discussed,  legitimate) arbitrariness of this choice, the statement regarding the exact Einstein nature of 
$\text{d}s^2_{\text{res.}}$ could hold. There is a simple explanation for this. 
 
Firstly, we have imposed (as part of our resummation ansatz) $\text{u}$ to be a shearless congruence. 
This assumption, not only enables us to discard the large number of Weyl-covariant tensors available when 
the shear is non-vanishing, which would have probably spoiled any resummation attempt; but it also selects 
the algebraically special geometries, known to be related with integrability properties. Indeed, on the bulk 
\eqref{papaefgenresc}, $\text{u}$ is a manifestly null congruence, associated with the vector $\partial_r$\,. 
One can show (see \cite{Petropoulos:2015fba}) that this bulk congruence is also \emph{geodesic} and \emph{shear-free}. 
 According to the generalizations of the Goldberg--Sachs theorem, the anticipated Einstein bulk metric  
\eqref{papaefgenresc}  is therefore algebraically special, \emph{i.e.} of Petrov type II, III, D, N or O. 

Secondly, the freedom in choosing $\text{u}$ is only apparent because we have required it to be shearless.
In $2+1$ dimensions, such a time-like vector field is essentially unique -- unless there are symmetries, 
in which case all choices are anyway equivalent due to the symmetries. Indeed, given a generic three-dimensional 
metric (rather, a conformal class of metrics), there is a unique way to express it as a fibration over a conformally
 flat two-dimensional base:\footnote{See e.g. \cite{Coll} and the discussion in the appendix of \cite{Petropoulos:2015fba}.}
\begin{equation}
\label{PDbdymetc}
\text{d}s^2=-(\text{d}t-\text{b})^2+\frac{2}{k^2P^2}\text{d}\zeta\text{d}\bar\zeta,
\end{equation}
with $P$ an arbitrary real function of $(t,\zeta, \bar \zeta)$, 
and
\begin{equation}
\label{frame}
\text{b}=B(t,\zeta, \bar \zeta)\, \text{d}\zeta+\bar B(t,\zeta, \bar \zeta)\, \text{d}\bar\zeta.
\end{equation}
In this metric, 
\begin{equation}
\label{ut}
\text{u}= -\text{d}t+\text{b}
\end{equation}
is precisely normalized and shear-free (see \cite{Petropoulos:2015fba}).   This defines our fluid congruence.

Thirdly, using the above resummation technique, it is possible to control \emph{from the boundary} the 
Petrov type of the bulk, encoded in the Weyl tensor. The Weyl tensor and its dual can be used to form a 
pair of complex-conjugate tensors. Their five independent complex components are naturally packaged 
inside two complex-conjugate symmetric $3 \times 3$ matrices $\text{Q}^{\pm}$ with zero trace (see e.g. \cite{GP}). 
The eigenvalue structure of $\text{Q}^{\pm}$ (\emph{i.e.} the degeneracy of the Weyl principal null directions) 
determines the Petrov type.
Performing the Fefferman--Graham expansion of the complex Weyl tensors $\text{Q}^{\pm}$ for a general 
Einstein space, one can show  \cite{Mansi:2008br, Mansi:2008bs,  deHaro:2008gp, Olea}
that the leading-order ($\nicefrac{1}{r^3}$) coefficients $\text{S}^{\pm}$ are related to the combination 
\begin{equation}
\label{eqn:Tref}
	T_{\mu\nu}^\pm = T_{\mu\nu} \pm \frac{i}{8\pi G k^2 }C_{\mu\nu} 
\end{equation}
of the components of the boundary Cotton and energy--momentum tensors, by a constant similarity 
relation: $\text{T}^\pm =- \text{P} \, \text{S}^{\pm}\text{P}^{-1}$ with $\text{P}={\rm diag}(\pm i,-1,1)$.
The Segre type of $ \text{S}^{\pm}$ determines precisely the Petrov type of the four-dimensional 
bulk metric and establishes a one-to-one map between the bulk Petrov type and the boundary data.
 We will see more precisely how this operates in the case of Robinson--Trautman spacetime. Notice 
for the moment that due to conservation equations \eqref{C-cons} and  \eqref{T-cons}, 
\begin{equation}
\label{Tref-cons}
\nabla\cdot \text{T}^\pm=0.
    \end{equation}

It is clear from the above that the absence of shear for the boundary fluid congruence plays 
a crucial r\^ole in the resummability of the derivative expansion, leading ultimately to exact
 algebraically special Einstein spaces. Nonetheless, we  cannot exclude that some exact Einstein 
type I spaces might be successfully reconstructed, or that none exact resummation involves a 
congruence with shear. In favour of the first option, one could argue that, the velocity of a 
relativistic fluid being arbitrary, one can always choose it shearless, without loss of generality. 
However, the way this congruence enters the resummation formula suggests, via the Goldberg--Sachs theorem, 
that we can only reach algebraically special Einstein spaces. We see thus the importance of this congruence 
from the holographic viewpoint, since it crucially enters and characterizes the resummation process. It is 
the reason why we proceed in the next section with the hydrodynamic analysis based on this congruence, which 
turns out to describe the holographic fluid in the Eckart frame. 

\subsection{The reconstruction of Robinson--Trautman} \label{sec:RT}
    
Consider the boundary metric   
\begin{equation}
\label{RTbdymet}
\text{d}s^2=-\text{d}t^2+\frac{2}{k^2P^2}\text{d}\zeta\text{d}\bar\zeta.
\end{equation}
The vector $\partial_t$ is hypersurface-orthogonal, and the normal hypersurfaces are constant-$t$ sections. 
The Gaussian curvature of the latter is $k^2 K$, where 
\begin{equation}
K=\Delta\ln P
\end{equation}
with $\Delta = 2P^2 \partial_{\bar\zeta} \partial_\zeta$.
The Cotton tensor, computed using\footnote{Together with the choice of retarded time quoted in 
note \ref{rminusr}, we reverse here the orientation with respect to the one adopted in \cite{Gath:2015nxa}:
$\eta_{t\zeta\bar\zeta}=\frac{i}{k^2P^2}$. With these conventions, time flows as in \cite{Bakas:2014kfa}, but is reversed with respect to Ref. \cite{deFreitas:2014lia}. Incidentally, we also rescale some observables for convenience,
resulting e.g. in extra $\nicefrac{1}{k^2}$ factors, as 
in Eq. \eqref{upm}.}  \eqref{cotdef}, reads:
\begin{equation}
\label{cot}
\text{C}=i \left(\begin{matrix}
\text{d}t& \text{d}\zeta & \text{d}\bar\zeta
\end{matrix}\right)
\left(\begin{matrix}
 0&-\frac{k^2}{2}\partial_\zeta K&\frac{k^2}{2}\partial_{\bar\zeta} K\\
-\frac{k^2}{2}\partial_\zeta K &-\partial_t\left(\frac{\partial^2_\zeta P}{P}
\right) &0 \\
\frac{k^2}{2}\partial_{\bar\zeta} K&0&\partial_t\left(\frac{\partial^2_{\bar\zeta} P}{P}\right) 
 \end{matrix}\right) \left(\begin{matrix}
\text{d}t \\ \text{d}\zeta \\ \text{d}\bar\zeta
\end{matrix}\right),
\end{equation}
which is a real tensor.

We must now introduce the canonical \emph{reference} tensors $\text{T}^\pm$ and apply the following strategy
 (valid more generally \emph{i.e.} beyond the choice \eqref{RTbdymet} of boundary metric):
\begin{enumerate}
\item  \label{step1} Determine the components of $\text{T}^\pm$ in terms of third derivatives of the boundary metric 
\eqref{RTbdymet}, using Eq. \eqref{cot} in (see \eqref{eqn:Tref})
\begin{equation}
\label{C-con}
 \text{Im} \text{T}^+=\frac{\text{C}}{8\pi G k^2}.
  \end{equation}  
\item  Use this information for expressing the actual energy--momentum tensor
\begin{equation}
\label{T-con}
\text{T}= \text{Re} \text{T}^+
    \end{equation}
in terms of third derivatives of the metric.   
\item Reconstruct the bulk spacetime metric using \eqref{papaefgenresc}. 
\item   \label{step3} Impose the conservation of $\text{T}$ \eqref{T-cons} and obtain a set of three \emph{a priori}
 fourth-order partial-differential equations for the boundary metric, which 
\begin{enumerate}
\item play the r\^ole of resummability conditions for the derivative expansion,
\item capture the boundary fluid dynamics.
\end{enumerate}
\end{enumerate}
Several remarks are in order here. First, the partial-differential equations obtained in step number \ref{step3}
 guarantee that Einstein's equations are fulfilled with the resummed derivative expansion \eqref{papaefgenresc}.
 Second, in step number \ref{step1}, Eq. \eqref{C-con} may impose restrictions among the components of the metric
 (its third derivatives in fact). These, with whatever external further condition we may impose via the form of 
$\text{T}^\pm$, control the Petrov type of the bulk. 

The power of the method displayed here is that we do not make any ansatz for the form of the energy--momentum
 tensor $\text{T}$. Rather we supply the reference tensors $\text{T}^\pm$ with a canonical form, which in 
turn delivers $\text{C}$ and $\text{T}$. The latter leads to equations for the boundary metric, which are 
also the holographic fluid equations of motion. 

Notice that we have no control on the frame in which the fluid is described, as the velocity field is the
shearless congruence read off directly from the boundary metric \eqref{RTbdymet}
 (see \eqref{ut}):
 \begin{equation}
 \label{uphys}
 \text{u}=-\text{d}t, 
  \end{equation}
which has no vorticity, no acceleration but is expanding at a rate
\begin{equation}
\label{expa}
\Theta=-2 \partial_t \ln P.
\end{equation}
 We should already stress that in this frame, which we will describe more precisely later, the holographic fluid 
exhibits a finite number of corrections with respect to a perfect fluid, as the energy--momentum tensor is basically 
third-order in derivatives of geometric quantities. This is not surprising 
and it is a rather general feature of exact Einstein bulk spaces to lead to holographic fluid configurations 
which do not trigger all transport coefficients. Still, the kinematic state is non-trivial, and the absence of 
certain series of corrections in the energy--momentum tensor is really the signature of vanishing of the 
corresponding transport coefficients (see \cite{Mukhopadhyay:2013gja} for the original  detailed discussion).
    
There are two basic and distinct canonical forms for $\text{T}^\pm$, which exhaust all possibilities. 
\begin{description}
\item[Perfect-fluid form] For perfect-fluid reference tensors, we need two complex-conjugate reference
 velocity fields $\text{u}^{\pm}$. Consider the normalized congruence\footnote{This is the most general one: 
adding an extra leg along the missing direction, and adjusting the overall scale for keeping the norm to $-1$ 
amounts to the combination of a Weyl transformation and a diffeomorphism.}
 \begin{equation}
 \label{upm}
 \text{u}^+= \text{u}+ \frac{\alpha^+}{k^2P^2}\text{d}\zeta
  \end{equation}
with
$\alpha^+= \alpha^+(t,\zeta,\bar \zeta)$, and its complex-conjugate
 $\text{u}^-= \text{u}+ \frac{\alpha^-}{k^2P^2}\text{d}\bar \zeta$ with $\alpha^-=\alpha^{+\ast}$. 
The perfect-fluid energy--momentum tensors based on these reference congruences read:
\begin{equation}
\label{RT-perflu}
\text{T}^\pm_{\text{pf}}= \frac{M_\pm(t,\zeta,\bar \zeta) k^2}{8\pi G}\left(3\left(\text{u}^\pm\right)^2 +\text{d}s^2\right)
\end{equation}
with $M_-=M_+^\ast$.
\item[Radiation-matter form] 
Consider finally
\begin{equation}
\label{rmt}
\text{T}^{+}_{\text{rm}}= \frac{1}{4\pi G}
\text{d}\zeta\left(\beta \text{d}t+\frac{\gamma}{k^2} \text{d}\zeta\right).
\end{equation}
In this expression $\beta$ and $\gamma$ are \emph{a priori} functions of $t,\zeta$ and $\bar \zeta$. 
The tensor is the symmetrized  direct product of a light-like by a time-like vector. Notice that for 
vanishing $\beta$, we obtain a \emph{pure-radiation} tensor \emph{i.e.} the square of a null vector. 

\end{description}
    
We will consider a general reference tensor of the form
\begin{equation}
\label{gent}
\text{T}^{+}=
\text{T}^{+}_{\text{pf}}+\text{T}^{+}_{\text{rm}},
\end{equation}
the two components being given in Eqs. \eqref{RT-perflu} and \eqref{rmt}. For this combination, 
\begin{eqnarray}
\label{T-com}
&&8\pi G i\,   \text{Im}\text{T}^+=\nonumber \\&&  \left(\begin{matrix}
\text{d}t& \text{d}\zeta & \text{d}\bar\zeta
\end{matrix}\right)
\left(\begin{matrix}
\displaystyle{k^2\left(M_+-M_-\right)}
 &-\frac{3M_+\alpha^+}{2P^2} +\frac{\beta}{2}
 &\frac{3M_-\alpha^-}{2P^2}-\frac{\bar \beta}{2}
 \\
-\frac{3M_+\alpha^+}{2P^2} +\frac{\beta}{2}
&\frac{3M_+(\alpha^+)^2}{2P^4k^2}+\frac{\gamma}{k^2}&\frac{M_+-M_-}{2P^2}
\\
\frac{3M_-\alpha^-}{2P^2}-\frac{\bar \beta}{2}&\frac{M_+-M_-}{2P^2}&
-\frac{3M_-(\alpha^-)^2}{2P^4k^2}-\frac{\bar\gamma}{k^2}
\end{matrix}\right) \left(\begin{matrix}
\text{d}t \\ \text{d}\zeta \\ \text{d}\bar\zeta
\end{matrix}\right),
    \end{eqnarray}
while   
 \begin{eqnarray}
\label{RT-com}
&&8\pi G \,  \text{Re}\text{T}^+=\nonumber \\&&  \left(\begin{matrix}
\text{d}t& \text{d}\zeta & \text{d}\bar\zeta
\end{matrix}\right)
\left(\begin{matrix}
\displaystyle{k^2\left(M_++M_-\right)}
 &-\frac{3M_+\alpha^+}{2P^2} +\frac{\beta}{2}
 &-\frac{3M_-\alpha^-}{2P^2}+\frac{\bar \beta}{2}
 \\
-\frac{3M_+\alpha^+}{2P^2} +\frac{\beta}{2}
&\frac{3M_+(\alpha^+)^2}{2P^4k^2}+\frac{\gamma}{k^2}&\frac{M_++M_-}{2P^2}
\\
-\frac{3M_-\alpha^-}{2P^2}+\frac{\bar \beta}{2}&\frac{M_++M_-}{2P^2}&
\frac{3M_-(\alpha^-)^2}{2P^4k^2}+\frac{\bar\gamma}{k^2}
\end{matrix}\right) \left(\begin{matrix}
\text{d}t \\ \text{d}\zeta \\ \text{d}\bar\zeta
\end{matrix}\right).    
\end{eqnarray}
 
 The reference tensor at hand depends on four complex arbitrary functions of 
$t,\zeta$ and $\bar \zeta$: $M_+, \alpha^+,\beta$ and $\gamma$. We can now require 
\eqref{C-con}, using \eqref{cot} and \eqref{T-com}. The first observation is that 
this identification of the Cotton tensor demands 
\begin{equation}
\label{bondi}
M_+(t,\zeta, \bar \zeta)=M_-(t,\zeta, \bar \zeta),
\end{equation}
which we will name $M(t,\zeta, \bar \zeta)$, a real function. Furthermore,  it appears 
a pair of independent conditions plus their complex-conjugates. The first reads:
\begin{equation}
3M\frac{\alpha^+}{P^2}+\partial_{\zeta} K=
\beta
\quad \text{and\quad c.c. },\label{h2}
\end{equation}
while the second is 
\begin{equation}
\frac{3}{2}M\frac{(\alpha^+)^2}{P^4}+\gamma=\partial_t\left(\frac{\partial^2_{\zeta} P}{P}\right)
\quad \text{and\quad c.c. }.\label{h3}
\end{equation}

Equations \eqref{h2} and \eqref{h3} are \emph{algebraic} for the functions  $\alpha^\pm(t,\zeta,\bar \zeta)$,
 $\beta(t,\zeta,\bar \zeta)$ and  $\gamma(t,\zeta, \bar \zeta)$,  as well as the complex conjugate functions 
 $\bar\beta(t,\zeta,\bar \zeta)$ and  $\bar\gamma(t,\zeta, \bar \zeta)$. Extracting  these functions and inserting
 them back into  \eqref{RT-com}, 
we determine  using  \eqref{T-con} the boundary energy--momentum tensor in terms of third derivatives of 
the metric, as already anticipated:
\begin{equation}
\label{RT-full}
\text{T}=\frac{1}{16\pi G} \left(\begin{matrix}
\text{d}t& \text{d}\zeta & \text{d}\bar\zeta
\end{matrix}\right)
\left(\begin{matrix}
 4Mk^2&\partial_\zeta K&\partial_{\bar\zeta} K\\
\partial_\zeta K &\frac{2}{k^2}\partial_t\left(\frac{\partial^2_{\zeta} P}{P}\right)&\frac{2M}{P^2}
\\
\partial_{\bar\zeta} 
K&\frac{2M}{P^2}&\frac{2}{k^2}\partial_t\left(\frac{\partial^2_{\bar\zeta} P}{P}\right) \end{matrix}\right)
 \left(\begin{matrix}
\text{d}t \\ \text{d}\zeta \\ \text{d}\bar\zeta
\end{matrix}\right).
\end{equation}

We are now ready to proceed and write the bulk metric as obtained using the resummed version of the derivative 
expansion, Eq. \eqref{papaefgenresc}. We find:
\begin{equation}
\label{papaefgentetr}
\text{d}s^2_{\text{res.}} =-2\text{d}t(\text{d}r +H \text{d}t)+2\frac{r^2}{P^2}\, \text{d}\zeta \text{d}\bar\zeta
\end{equation}
with
\begin{equation}
\label{H}
2H = k^2 r^2 - 2r \partial_t \ln P+  K -\frac{2M}{r}\,.
\end{equation}
According to our reasoning about the resummation of the derivative expansion into an exact Einstein space, 
the metric \eqref{papaefgentetr} is expected to be Einstein provided the boundary energy--momentum tensor 
\eqref{RT-full} is conserved, \emph{i.e.} obeys \eqref{T-cons}. Let us impose therefore the conservation of $\text{T}$:
\begin{equation}
\label{E}
\nabla\cdot \text{T}=0\quad\Longleftrightarrow\quad
\begin{cases} \Delta K +12M \partial_t \ln P=4  \partial_t M,\\
\partial_\zeta M=0,\quad
\partial_{\bar\zeta} M=0.
\end{cases}
    \end{equation}
Not only the first equation in \eqref{E} is the \emph{Robinson--Trautman} equation, which precisely guarantees that 
 \eqref{papaefgentetr} is Einstein, but it also appears here as the longitudinal component of the energy--momentum 
conservation, \emph{i.e.} as the \emph{heat} equation for the boundary fluid, at rest in the frame at hand. We will
 further elaborate on the properties of the holographic fluid in the next section. 

We would like at this point to remark that no reference to any \emph{a priori} bulk property has been made in our
 approach. The Robinson--Trautman equation has been obtained from purely boundary considerations, by imposing the 
conservation of the boundary energy--momentum tensor, and we can similarly tune the boundary data in order to control
 the bulk Petrov type of the bulk Einstein space. Generically the latter is type II because we can prove
 \cite{Petropoulos:2015fba} that the bulk congruence $\partial_r$ is null, geodesic and shearless, and using 
thus the extensions of Goldberg--Sachs theorem, the reconstructed bulk space is algebraically special.\footnote{Notice that 
Robinson--Trautman spacetimes were originally designed to be algebraically special -- see \cite{Podolsky:2016sff}
 for more information regarding the principal null directions of Robinson--Trautman.} By tuning the functions that
 define the reference tensors $\text{T}^\pm$, namely
$M(t), \alpha^\pm(t,\zeta,\bar \zeta)$, $\beta(t,\zeta,\bar \zeta)$, $\bar\beta(t,\zeta,\bar \zeta)$,
 $\gamma(t,\zeta, \bar \zeta)$  and  $\bar\gamma(t,\zeta,\bar \zeta)$, we can scan other 
classes (see \cite{Gath:2015nxa} for details):
\begin{itemize}
\item If $M=0$, $\alpha^\pm$ are immaterial and
$\beta(t,\zeta,\bar \zeta)$ and $\gamma(t,\zeta, \bar \zeta)$ are fully determined by Eqs. \eqref{h2} and  \eqref{h3}: 
\begin{eqnarray}
\beta&=&\partial_{\zeta} K\quad \text{and\quad c.c. },\label{h2Mzero}\\
\gamma&=&\partial_t\left(\frac{\partial^2_{\zeta} P}{P}\right)\quad \text{and\quad c.c. }.\label{h3Mzero}
\end{eqnarray}
Furthermore, the Robinson--Trautman equation guarantees holomorphicity for $\beta$, function of $(t,\zeta)$ only.
Hence, the bulk is generically Petrov type III. When $\beta=0$, it becomes type N, where now $K=K(t)$, following \eqref{h2Mzero}. 
The most general $P(t,\zeta,\bar \zeta)$ such that its curvature is a function of time
 only was found in \cite{Foster:1967:NRT}, and reads:
\begin{equation}
P(t,\zeta,\bar \zeta)=\frac{1+\frac{\epsilon}{2} h(t,\zeta)\, \bar h(t,\bar\zeta)}{\sqrt{2f(t)\, 
\partial_\zeta h(t,\bar\zeta)\, \partial_{\bar\zeta}\bar h(t,\bar\zeta)}}
\end{equation}
with $\epsilon=0,\pm1$ and arbitrary functions $f(t)$ and $h(t,\zeta)$.
 \item If $\beta=\gamma=0$, $\alpha^\pm$ are read-off from \eqref{h2}:
 \begin{equation}
\alpha^+=-\frac{P^2}{3M}\partial_{\zeta} K\quad \text{and\quad c.c. },\label{h2bczero}
\end{equation}
and the geometry is subject to a further constraint\footnote{Notice a useful identity:
$\partial_t\left(\frac{\partial^2_{\zeta} P}{P}\right) =
\frac{1}{P^2}\partial_{\zeta} \left(P^2\partial_t\partial_{\zeta}
\ln P
\right)$.\label{identity}} obtained by combining \eqref{h3} and \eqref{h2bczero}:
\begin{equation}
6M\,\partial_t\left(\frac{\partial^2_{\zeta} P}{P}\right) =\left(\partial_{\zeta} K\right)^2
\quad \text{and\quad c.c. }.\label{h3bczero}
\end{equation}
The bulk is still type II, but choosing holomorphic $\alpha^-=\alpha^-(t,\zeta)$, \emph{i.e.} (using \eqref{h2bczero})
\begin{equation}
\partial_{\zeta}\left(P^2 \partial_{\zeta} K\right)=0\quad \text{and\quad c.c. },\label{h2bczeroholo}
\end{equation}
together with the constraint \eqref{h3bczero}, makes it type D.  There are two independent type D solutions:
\begin{enumerate}
\item The Schwarzschild, reached with $P=1+\frac{\epsilon}{2}\zeta\bar \zeta$ and $K=\epsilon$, which 
is asymptotically anti-de Sitter.
\item The $C$-metric, which requires $P^2 \partial_{\zeta} K=h(\bar \zeta)\neq 0$ and is asymptotically 
locally anti-de Sitter due to a non-vanishing boundary Cotton tensor.
\end{enumerate}
\end{itemize}

Let us mention here that the time dependence of $M$ remains arbitrary, and can be reabsorbed by performing 
 an appropriate bulk diffeomorphism, inducing a conformal transformation plus a diffeomorphism on the boundary 
\cite{GP}. The Robinson--Trautman equation reads then: 
\begin{equation}
\label{RT-time-ind}
 \partial_{\bar\zeta} \partial_\zeta K =3M \partial_t\left( \frac{1}{P^2}\right)
\end{equation}
with constant $M$. We will adopt this convention for the rest of our presentation.

Before moving to the hydrodynamic analysis of the energy--momentum tensor, we would like to end the current section 
with some general comments regarding the bulk Einstein spaces under consideration. 

With the exception of the Petrov-D solutions quoted above, Robinson--Trautman spacetimes are time-dependent and 
carry gravitational radiation. Once this radiation is emitted, the spacetime settles down generically to an 
anti-de Sitter Schwarzschild black hole.  The general features of this evolution are captured by the Robinson--Trautman
 equation, which, following \cite{Tod:1989}, is a parabolic equation describing a Calabi flow on a two-surface.  
As long as $M\neq 0$, these spacetimes exhibit a past singularity at $r=0$, past-trapped two-surfaces and a future
 horizon, which is the anti-de Sitter Schwarzschild horizon at late times. Unfortunately, singularities are often 
developed on this horizon and no smooth extension is possible beyond, in the interior region. 

Irregularities of the two-surface $\mathcal{S}$ time-dependent metric 
\begin{equation}
\label{RTbdymetspace}
\text{d}\ell^2=\frac{2}{k^2P(t,\zeta,\bar\zeta)^2}\text{d}\zeta\text{d}\bar\zeta,
\end{equation}
possibly present at early times, are washed out by the evolution, as usual with geometric flows. The flow at hand,
 governed by the Robinson--Trautman equation \eqref{RT-time-ind}, has the following salient properties: 
 \begin{eqnarray}
\label{RTproperties}
&&\frac{\text{d}}{\text{d}t}\int_\mathcal{S}\frac{\text{d}^2\zeta}{P^2}=0,\\
\label{RTproperties2}
&&\frac{\text{d}}{\text{d}t}\int_\mathcal{S}\frac{\text{d}^2\zeta}{P^2}K=
0,
\end{eqnarray}
where $\text{d}^2\zeta=-i\,\text{d}\zeta\wedge\text{d}\bar\zeta$ (this assumes there are no boundary-like contributions --
the proof will be given and commented in Sec. \ref{sec:RTfluid}). Hence,
the area of $\mathcal{S}$ and its average curvature (\emph{i.e.} the Euler number) are preserved along 
the flow, which, at late times, brings the metric into a symmetric geometry compatible with the original topology. 
From the spacetime perspective, this situation corresponds indeed to the evolution towards an anti-de Sitter Schwarzschild black 
hole with conformal boundary $\mathbb{R}\times S^2$, $E_2$  or $H_2$.\footnote{The Calabi flow is set for a metric 
on a compact K\"ahler space, here two-dimensional. For this reason it was quoted in \cite{Bakas:2014kfa} for
 spherical geometry only. Probably, $E_2$ or $H_2$ could also support this flow, assuming they were made 
compact by modding out some discrete isometry. This line has not attracted much attention, and at present 
Calabi-flow results do not cover all Robinson--Trautman geometries. The statements regarding late-time 
behaviour should therefore be taken with care as they have not been demonstrated for all possible initial conditions. 
In particular, the possibility of reaching the $C$-metric has been discussed in \cite{C-limit}. 
In that work it was shown that Robinson--Trautman spacetimes admitting a space-like isometry generically decay
to the C-metric.
\label{f8}} 

Closing this chapter, one should observe that Robinson--Trautman spacetimes appear as laboratories for 
investigating time-dependent black-hole exact solutions surrounded by gravitational radiation. As opposed 
to the stationary paradigms, very little is known here, even at a very elementary level: location of past
 horizon, definition of thermodynamic quantities such as energy, temperature or entropy, interpretation of
 the evolution as out-of-equilibrium thermodynamics. This is surprising because understanding deviations from 
equilibrium in these systems is at least as important as counting their microscopic degrees of freedom, which 
has attracted more attention. Any further comment on bulk thermodynamics  would be, at this stage, daring.

\section{The  Robinson--Trautman holographic fluid}\label{sec:RTfluid}

Following the general plan presented in Sec. \ref{sec:resform},  we have reached Robinson--Trautman spacetimes
 in Sec. \ref{sec:RT}, using in the derivative expansion \eqref{papaefgenresc}, the boundary metric \eqref{RTbdymet},
 the boundary energy--momentum tensor \eqref{RT-full} and the boundary fluid velocity field $\eqref{uphys}$. The latter
 defines the hydrodynamic frame where the resummation of the derivative expansion is successfully performed -- for reasons that we 
have already discussed. This frame turns out to be very natural  for describing the fluid properties. 
 
 \subsection{The hydrodynamic frame and the fluid transport data}

In the case at hand, the energy density of the fluid reads:\footnote{As pointed out in App. \ref{out}, 
the kinematical out-of-equilibrium quantities $\pmb{\varepsilon}$, $\pmb{p}$ and $\pmb{\varrho}$ are chosen 
to coincide with the thermodynamic local-equilibrium $\varepsilon$, $p$ and $\varrho$.}
\begin{equation}
\label{endenRT}
\varepsilon = T_{\mu\nu}u^\mu u^\nu= \frac{M k^2}{4\pi G},
\end{equation}
and is constant, as is the pressure ($\varepsilon=2p$). 
We can split  the energy--momentum tensor (see App. \ref{cong} and e.g. \cite{Kovtun:2012rj, Romatschke:2009im}) as 
\begin{equation}
\label{Tdec}
T_{\mu\nu}=T^{(0)}_{\mu\nu}+\tau_{\mu\nu}   +u_\mu  q_\nu  + u_\nu q_\mu ,
\end{equation}
with a conformal-perfect-fluid part
\begin{equation}
\label{Tperf}
\text{T}^{(0)}=\frac{\varepsilon}{2}\left(3\text{u}^2+\text{d}s^2\right) 
\end{equation}
and a non-perfect piece $\tau_{\mu\nu}   +u_\mu  q_\nu  + u_\nu q_\mu$, where
 $\tau_{\mu\nu}$ and $q_\mu$ are the components of the \emph{stress tensor} and the \emph{heat current} respectively. 
These are fully transverse:
\begin{equation}
\label{pitra}
\tau_{\mu\nu}u^\mu = 0, \quad q_\mu u^\mu=0
\end{equation}
with
\begin{equation}
\label{qdef}
 q_\nu= -\varepsilon u_\nu
 -u^\mu  T_{\mu \nu}.
 \end{equation}
The non-perfect piece $u_\mu  q_\nu  + u_\nu q_\mu$
is  \emph{non-transverse}. The latter is absent in the Landau--Lifshitz frame.

Here we are \emph{not in the Landau--Lifshitz}, but rather \emph{in the Eckart} frame (see App. \ref{out} for a detailed discussion on this subject). To show this we should 
consider the more general charged Robinson--Trautman solution, which solves bulk Einstein--Maxwell
 equations and has a conserved current $\text{J}$ on the boundary.\footnote{Conserved currents may also appear without extra degrees of freedom, in systems with symmetries generated by Killing vectors $\text{k}$. Indeed, in those situations $k_\nu T^{\mu\nu}$ are components of divergence-free vectors. Since Robinson--Trautman spacetimes have generically no isometries, we will not investigate this direction.} In these solutions, the electromagnetic field has 
three components: magnetic,  electric and radiation.  On the boundary, there is a conserved current, a chemical 
potential and a magnetic field \cite{CM}. The latter couples to the current as 
$\nabla_\mu T^{\mu\nu}=4\pi G J_\mu F^{\mu\nu}$, and vanishes if and only if the bulk 
radiation component is absent. In this case of ideal magnetohydrodynamics,\footnote{Keeping the radiation component opens the field of general magnetohydrodynamics -- see \cite{Caldarelli} for a related discussion, and \cite{Kovtun} for a more general perspective.} is again governed by the plain Robinson--Trautman equation, and the conserved current has the perfect form ($j_\nu=0$ in \eqref{curdec}):
\begin{equation}
\label{Jdef}
 J_\nu= \varrho u_\nu
 \end{equation}
with
\begin{equation}
\label{rhodef}
 \varrho =\frac{k^2 Q}{4\pi G} P(t,\zeta,\bar \zeta)^2
 \end{equation}
 and $Q$ an arbitrary constant.
This demonstrates the statement regarding the Eckart frame, since the current is fully longitudinal and perfect. 

In the Eckart frame, the heat current is non-vanishing and we find, using \eqref{qdef},
\begin{equation}
\label{q}
\text{q} = -\frac{1}{16\pi G}\left(\partial_\zeta K \, \text{d}\zeta+ \partial_{\bar\zeta}\,  K\text{d}\bar\zeta \right).
\end{equation}
The non-perfect stress tensor (we have used the identity of footnote \ref{identity}) is given by
\begin{equation}
\label{pi}
\tau  = \frac{1}{8\pi Gk^2P^2}
\left(
\partial_{\zeta} \left(P^2\partial_t\partial_{\zeta}
\ln P
\right)\text{d}\zeta^2+ 
\partial_{\bar\zeta} \left(P^2\partial_t\partial_{\bar\zeta}
\ln P
\right)
\text{d}\bar\zeta^2 \right).
\end{equation}
It reflects the friction, which is of kinematic origin. Hence, it is not surprising that we can express it in terms of
 the orthogonally projected covariant derivatives (see App. \ref{cong}) of the fluid velocity:\footnote{In our case,
 due to the absence of shear, vorticity and acceleration, the velocity derivatives are expressed only in terms of 
derivatives of the expansion, as for example:
$$
\nabla_{\lambda}\nabla_{\mu}u_{\nu}=
\frac{1}{2}\partial_{\lambda}\Theta h_{\mu\nu}+\frac{1}{4}\Theta^2\left(h_{\lambda \mu}u_{\nu}+h_{\lambda\nu}u_{\mu}\right)\,.
$$}
\begin{eqnarray}
\nonumber
\tau_{\mu\nu}&=& -\frac{1}{16\pi Gk^2}\left(
D_\mu D_\nu \Theta -\frac{1}{2} h_{\mu\nu} D^\lambda D_\lambda \Theta \right)\\
\label{Sperp}
&=&-\frac{1}{16\pi Gk^2}\left(h_\mu^{\hphantom{\mu }\rho} h_\nu^{\hphantom{\nu }\sigma} 
\nabla_\rho h_\sigma^{\hphantom{\sigma }\lambda}\nabla_\lambda \Theta -\frac{1}{2} h_{\mu\nu} \nabla_\rho h^{\rho\sigma}
\nabla_\sigma \Theta \right).
\end{eqnarray}
This is not possible for $\text{q}$ though. Generically, the heat flow cannot be expressed as a pure $\text{u}$-derivative
 expansion, it also involves the gradient of scalars like the temperature or the curvature, and betrays   thermal conduction
 or similar phenomena.

As already mentioned in Sec. \ref{sec:RT}, when dealing with exact algebraically special Einstein spaces, the holographic
 energy--momentum tensor receives at most third-order derivative corrections with respect to the perfect fluid. The reason
 is simple. The bulk algebraic structure sets an intimate relationship between the energy--momentum tensor and the Cotton 
tensor, which is a third derivative of the boundary metric. Since the shearless velocity field is determined by the
 geometry itself, the energy--momentum is necessarily expressed with third derivatives of the  velocity field.

This property is very general. It was extensively discussed in a wide class of situations like the Pleba\'nski--Demia\'nski family, where the energy--momentum tensor is either third-order in 
$\text{u}$-derivatives (in the presence of a bulk acceleration parameter) \cite{Petropoulos:2015fba}, or is perfect \cite{Mukhopadhyay:2013gja}. 
This latter case does not imply that the fluid is perfect: some of the would-be corrections vanish just because of kinematic reasons (as $-2\eta \sigma_{\mu\nu}$), some other  because infinite series of transport coefficients are indeed zero for 
the holographic fluid at hand. 

In the Robinson--Trautman case, the unique available transport coefficient is read-off
in  $\text{q}$ (Eq. \eqref{q}) or in $\tau$ (Eq. \eqref{pi}). This coefficient is of order $\nicefrac{1}{16\pi G}$, and we will further comment on it in Sec. \ref{eckart}. As long as we remain within Robinson--Trautman solutions, this is the only information we can get, and it is exact. Of course, in order to 
have access to more  transport coefficients (possibly infinite series of them),
we can consider changing hydrodynamic frame. But even in that case, the new ones will all stem out of the former, and all will be of the same order. 

For example, it is possible to move from Eckart to Landau--Lifshitz frame. As explained thoroughly in App. \ref{out}, this requires some care. At the first place, these frames are built assuming the existence of a conserved matter current. Moving from Eckart to Landau--Lifshitz trades the heat current of the conserved energy--momentum tensor in Eckart for the transverse part of the matter current in Landau--Lifshitz. This is conceivable for the charged Robinson--Trautman, but audacious for the neutral case. At a second stage, the actual transformation is performed perturbatively, order by order in a parameter, which is $\|\text{q}\|$ (see App.~\ref{out} for detailed expressions), required to be small compared to the energy scale. These series are usually asymptotic.

This philosophy was originally pursued in \cite{deFreitas:2014lia}  with success regarding the determination of transport coefficients. 
Still, it has some caveats. From the mathematical viewpoint, this amounts to trading an exact quantity like 
$\tau$ or $\text{q}$, for an infinite series, which in general lacks convergence.
Physics-wise, moving to Landau--Lifshitz blurs the simple and clear picture, which emerges  in the Eckart frame as we will see; moreover, doing so while ignoring the matter current $\text{j}$ is
 inappropriate, in particular when computing the entropy current (see Sec.~\ref{entropycurrent}).\footnote{The same attitude was adopted later on by the authors of \cite{Bakas:2014kfa}, who insist in moving to Landau--Lifshitz 
in their follow-ups 
\cite{Bakas:2015hdc,Skenderis:2017dnh}.}

 \subsection{Physics and evolution in the Eckart frame} \label{eckart}

In the Eckart frame, the pressure is constant and the fluid is at rest on a spatial section $\mathcal{S}$ equipped with 
a metric $\text{d}\ell^2$ (Eq. \eqref{RTbdymetspace}). The physical phenomena taking place in the fluid are related to 
thermal conduction, materialized in the heat current $\text{q}$, Eq. \eqref{q}, and captured by the Robinson--Trautman 
equation \eqref{RT-time-ind} appearing as the time component of the energy--momentum conservation \eqref{E}. This is 
a heat-flow equation, and one can elegantly derive it directly from the general  heat-current-divergence equation 
displayed in \eqref{divq}. 
In the case under investigation, $a_\mu$, $\sigma_{\mu\nu}$ and $g_{\mu\nu} \tau^{\mu\nu}$ vanish, whereas $\varepsilon$ is 
constant, 
so \eqref{divq} reads:
\begin{equation}
\label{divq2}
\text{div}_{(2)}\text{q}=-\frac{3\varepsilon}{2}\Theta.
\end{equation} 
We have introduced $\text{div}_{(2)}\text{q}= \nabla_{(2)i}\, q^i$, which is equal to $\nabla_\mu q^\mu$ because $\text{q}$ is
 transverse with respect 
to the hypersurface-orthogonal vector $\text{u}=\partial_t$, so exclusively defined inside the spatial section $\mathcal{S}$.
 Geometric quantities 
referring to this surface and to the corresponding metric $\text{d}\ell^2$ will carry a subindex ``(2)'':
\begin{itemize}
\item antisymmetric tensor: $\eta_{(2)\zeta\bar\zeta}=-\frac{i}{k^2P^2}$, and volume 
form: $\Omega_{(2)}=-i\frac{\text{d}\zeta\wedge\text{d}\bar\zeta}{k^2P^2}=\frac{\text{d}^2\zeta}{k^2P^2}$;
\item Laplacian operator: $\triangle_{(2)}f=k^2 \triangle f=2k^2P^2\partial_\zeta\partial_{\bar\zeta}f$, 
and scalar curvature: $R_{(2)}=2 k^2 K$;
\item Hodge--Poincar\' e duality: $\text{q}=q_{\zeta} \text{d}\zeta+q_{\bar\zeta} 
\text{d}\bar\zeta \Leftrightarrow \underset{(2)}{\star}\text{q}=
i\left(q_{\zeta} \text{d}\zeta-q_{\bar\zeta} \text{d}\bar\zeta\right) $.
\end{itemize}
Substituting in Eq. \eqref{divq2} the heat current \eqref{q} expressed as
\begin{equation}
\label{q2}
\text{q} = -\frac{1}{16\pi G} \text{d}_{(2)}K,
\end{equation}
the expansion $\Theta$ given in \eqref{expa}, and the constant energy density \eqref{endenRT}, we find indeed the 
Robinson--Trautman equation \eqref{RT-time-ind}:
\begin{equation}
 \nonumber
 \partial_{\bar\zeta} \partial_\zeta K =3M \partial_t\left( \frac{1}{P^2}\right).
\end{equation}

Equation \eqref{divq2} can be used in integral form, over a fixed domain $\mathcal{D}\subseteq \mathcal{S}$ with 
boundary $\partial\mathcal{D}$. Thanks to Green's theorem,\footnote{Reminder of Green's theorem: for any vector/one-form $\text{v}$
$$
\int_\mathcal{D}\frac{\text{d}^2\zeta}{k^2P^2} \text{div}_{(2)}\text{v}
=\oint_{\partial\mathcal{D}} \underset{(2)}{\star}\text{v}.
$$} we find:
\begin{equation}
\label{divq2int}
\int_\mathcal{D}\frac{\text{d}^2\zeta}{k^2P^2} \varepsilon\, \Theta
=-\frac{2}{3}\oint_{\partial\mathcal{D}} \underset{(2)}{\star}\text{q}.
\end{equation}
Using specifically \eqref{expa} for $\Theta$, \eqref{endenRT} for $\varepsilon$ and \eqref{q2} for $\text{q}$, we finally obtain:  
\begin{equation}
\label{arealaw}
k^2\frac{\text{d}A_\mathcal{D}}{\text{d}t}=
\frac{i}{6M}\oint_{\partial\mathcal{D}} \left(\partial_\zeta K\text{d}\zeta-
\partial_{\bar \zeta}K\text{d}\bar \zeta\right),
\end{equation} 
where
\begin{equation}
A_\mathcal{D}= 
\int_\mathcal{D}\frac{\text{d}^2\zeta}{k^2P^2} 
\end{equation} 
is the area of the domain $\mathcal{D}$. Multiplying by $\varepsilon$, the total energy stored by the fluid inside $\mathcal{D}$, 
\begin{equation}
E_\mathcal{D}=   \frac{M}{4\pi G}\int_\mathcal{D}\frac{\text{d}^2\zeta}{P^2}
\end{equation} 
obeys
\begin{equation}
\label{energylaw}
\frac{\text{d}E_\mathcal{D}}{\text{d}t}=
\frac{i}{24\pi G}\oint_{\partial\mathcal{D}} \left(\partial_\zeta K\text{d}\zeta-
\partial_{\bar \zeta}K\text{d}\bar \zeta\right).
\end{equation} 

Assuming $\mathcal{S}$ be a compact surface without boundaries, from Eq. \eqref{arealaw}, we conclude that the total area
 of $\mathcal{S}$, $A=A_\mathcal{S}$ remains constant in time.\footnote{Under appropriate assumptions for $K$ asymptotics,
 $\mathcal{S}$ could even be non-compact, and its area infinite.} This demonstrates \eqref{RTproperties}. Accordingly, the 
total energy
$E=E_\mathcal{S}=\varepsilon A$ is also conserved. Along time, the spatial section $\mathcal{S}$ hosting the fluid evolves
 and the fluid energy, conserved in total, moves from one region to another. With reasonable initial conditions,
 the system stabilizes at large times in a configuration with spatially constant $K$ (see discussion at the end 
of Sec. \ref{sec:RT}).

Summarizing, the Robinson--Trautman holographic fluid is at rest in the Eckart frame and is subject to thermal conduction, 
with energy exchanges operating according to the dynamics described above, and driven by the heat current \eqref{q}.

In order to simplify our discussion and fit within the framework of the the Robinson--Trautman spacetime built in Sec. \ref{sec:RT}, we will consider from now on vanishing chemical potential. This choice is holographically achievable \cite{CM}. {We could alternatively set the density to zero; all of our conclusions would hold in that case, but we find the former option more convenient.} Following \eqref{stefan} and \eqref{stefanorig}, we find for the conformal fluid at hand the temperature as related to the energy density by standard Stefan's law:
\begin{equation}
\label{stefan-3}
\varepsilon = \sigma T^3 = \frac{M k^2}{4\pi G}
\end{equation} 
with $\sigma= \frac{8\pi^2 G^2}{27 k^4}$. Hence the local-equilibrium thermodynamic temperature $T$ is constant.

The heat current of the Robinson--Trautman fluid  can be expressed, like for any fluid, as a derivative expansion in the temperature, and in geometric or kinematic tensors. In the present case, however, this current is known exactly, and contains a single term, 
that would appear at third order in the derivative expansion. The would-be first-order term, displayed in the generic 
expression \eqref{q1}, is absent here. In this expression, appears the local-thermodynamic-equilibrium temperature $T$, given in \eqref{stefan-3}, which is constant. Since 
the acceleration is vanishing, the first order does not contribute indeed. 

One may be puzzled at this stage, discussing thermal conduction without temperature gradients.  This attitude is probably
 too naive. As explained in App. \ref{out}, quantities like temperature or chemical potential lack a microscopic definition 
when out-of-equilibrium phenomena take place. Even though the  hydrodynamic hypothesis of local thermodynamic equilibrium may
 be justified, the local-equilibrium temperature $T(x)$ (in fact constant here) or chemical potential $\mu(x)$ (absent in our 
case) do not exhaust all available information, and more is captured in the kinematical, out-of-equilibrium functions 
$\pmb{T}(x)$ and $\pmb{\mu}(x)$. 

The origin of the transport phenomena witnessed here being in essence geometric, it is tempting, inspired by \eqref{q1}, to
 recast the exact expression of the current \eqref{q} as
\begin{equation}
\label{qf}
q_\mu = -\kappa D_\mu \pmb{T}
\end{equation} 
with
\begin{equation}
\label{kinT}
\kappa \pmb{T}(t,\zeta,\bar \zeta) = \kappa T+ \frac{1}{16\pi G}\left(K(t,\zeta,\bar \zeta) -\left\langle
K 
\right\rangle
\right).
\end{equation} 
The Gaussian curvature $K(t,\zeta,\bar \zeta)$ contributes thus to a kind of kinematical, out-of-equilibrium temperature $\pmb{T}(t,\zeta,\bar \zeta)$. It is naturally accompanied with a heat conductivity, read off as its coefficient in  \eqref{kinT}:
\begin{equation}
\label{kappa}
\kappa = \frac{1}{16\pi G}.
\end{equation} 
The latter is of geometric origin, as the transport phenomenon it triggers. This result is in agreement with the general analysis performed in \cite{Iqbal:2008by}.

In expression \eqref{kinT}, we have introduced $T$ given in \eqref{stefan-3}, and the average curvature\footnote{Defined as a limit for a non-compact surface.} over $\mathcal{S}$:
\begin{equation}
\langle
K 
\rangle= \frac{1}{A}
\int_\mathcal{S}\frac{\text{d}^2\zeta}{k^2P^2} K.
\end{equation} 
This turns out to be constant, as advertised in \eqref{RTproperties2}. Indeed, one easily shows that
\begin{equation}
\label{RTproperties2new}
\frac{\text{d}}{\text{d}t}\int_\mathcal{D}\frac{\text{d}^2\zeta}{P^2}K=
-\frac{i}{2}\oint_{\partial\mathcal{D}}
\left(\partial_{\zeta}\Theta\, \text{d}\zeta-
\partial_{\bar \zeta}\Theta\, \text{d}\bar \zeta\right),
\end{equation} 
which vanishes when $\mathcal{D}= \mathcal{S}$, under the already spelled assumptions.\footnote{The identity \eqref{RTproperties2new} does not require the Robinson--Trautman equation to be satisfied. It is thus valid for any dynamics and not necessarily for the Calabi flow. Actually it reads: 
$$
\frac{\text{d}}{\text{d}t}\int_\mathcal{D}\frac{\text{d}^2\zeta}{P^2}\triangle f=
i\oint_{\partial\mathcal{D}}
\left(\partial_{\zeta}\partial_t f\, \text{d}\zeta-
\partial_{\bar \zeta}\partial_t f\, \text{d}\bar \zeta\right),
$$
for any function $f(t,  \zeta, \bar \zeta)$.} 
For asymptotic time,  $K(t,\zeta,\bar \zeta)$ is expected to converge towards a constant, which is therefore identified with $\langle K \rangle$. Hence 
\begin{equation}
\label{Tlim}
\lim_{t\to+\infty}
\pmb{T}(t,\zeta,\bar \zeta) =T.
\end{equation} 
At late times, the fluid reaches global equilibrium with the kinematical temperature equal to the thermodynamic-equilibrium temperature, as expected. At any time, the thermodynamic-equilibrium temperature is the average kinematical temperature: 
$\langle
\pmb{T}(t,\zeta,\bar \zeta) 
\rangle =T$. 

The validity of holographic approach in the present framework requires a large black-hole mass, hence a large temperature $T$.
This leaves room for initial conditions on $P(t,\zeta,\bar \zeta)$ that do not violate the positivity of 
$\pmb{T}(t,\zeta,\bar \zeta)$. Actually, the latter may not be mandatory since 
$\pmb{T}(t,\zeta,\bar \zeta) $ is an instrument for probing transport, and not a fundamental quantity defined \emph{ab initio} -- reason why we insist calling it ``kinematical, out-of-equilibrium temperature''  as opposed to ``local-thermodynamic-equilibrium temperature'' (see discussion in App. \ref{out}). 

 \subsection{The entropy current and its conservation}\label{entropycurrent}

The last important aspect of the Robinson--Trautman fluid dynamics we would like to discuss is the entropy, the associated current and its divergence.   
For the conformal case in three dimensions, the standard entropy current is given in \eqref{curspERTcharged} in the Eckart frame,
and
reproduced here for clarity:
\begin{equation}
\label{curspERTcharged3}
S^\mu=\frac{1}{T}\big((3p-\mu\varrho) u^\mu+q^\mu\big).
\end{equation} 
We remind that in this expression the local-equilibrium thermodynamic quantities and relations are used, following the discussion of App. \ref{out}, as determined in the Eckart frame. It applies to the more general charged Robinson--Trautman solution with density displayed in Eq. \eqref{rhodef}. Since we have chosen zero chemical potential, the second term drops,\footnote{{This term also drops for vanishing density.}} and the entropy is constant:
\begin{equation}
\label{entroRT}
s=\frac{3p}{T} =\frac{3\sigma T^2}{2}=\left( \frac{M}{4}\right)^{\nicefrac{2}{3}} .
\end{equation} 
In this case, the entropy current reads:
\begin{equation}
\label{entrocurRT}
\text{S}=
\left( \frac{M}{4}\right)^{\nicefrac{2}{3}}
\left( \partial_t- \frac{P^2}{6M}
\left(
\partial_{\bar\zeta}K\,
\partial_\zeta
+
\partial_\zeta K\,  
\partial_{\bar\zeta}
\right)
\right).
\end{equation} 
Using the general expression for the entropy-current divergence \eqref{curspconf}, we obtain:
\begin{equation}
\label{divS}
\nabla_\mu S^\mu=0.
\end{equation} 
This is the consequence of the local-equilibrium temperature and pressure being constant, and of the vanishing chemical potential, shear and acceleration. Put differently, $s$ and $T$ being both constant,  
the current $\text{S}$ is divergence-free as a consequence of a fine cancellation between the velocity expansion $\Theta$ and the divergence of the heat current, displayed in \eqref{divq2}.

The conservation of the entropy current is surprising at first sight because we are seemingly out of equilibrium and evolution towards equilibrium usually produces entropy. However, the thermal-conduction irreversible phenomenon described by the Robinson--Trautman dynamics is of geometric nature. Hence, it can reasonably accommodate a conserved entropy current.
Indeed, the fluid is at rest. The evolution preserves the area and the energy, 
and occurs at a constant average kinematical temperature, equal to the local-equilibrium temperature. At the same time the absence of acceleration and shear wash out the effects of the heat current and the stress friction (see \eqref{curspEN}), and 
the process ultimately appears as an adiabatic, even isentropic, redistribution of energy {due to the kinetics of the surface rather than to the motion of the fluid}, till the final global-equilibrium state is reached. In thermodynamic language this is a special case of \emph{isothermal} Carnot's path,\footnote{We use intentionally ``path'' rather than ``cycle'' as in the process under consideration the system does not come back to the original state because of the time-evolving geometry.} known as Moutier's \cite{Moutier}, which produces no work and has zero thermodynamic efficiency.\footnote{The thermodynamic efficiency of a cycle is defined as $\pmb{\eta}=1-\nicefrac{T_{\text{min}}}{T_{\text{max}}}$.} Carnot's evolution is reversible and this does not contradict anything here, as the origin of irreversibility for the described phenomenon is purely geometrical.

The above conclusion is \emph{frame-independent} as is the actual entropy current. The latter can be expressed alternatively as in Eq. \eqref{curspL}:
\begin{equation}
\label{curspLRT}
\text{S}=
s_{\text{LL}} \text{u}_{\text{LL}}-\frac{\mu_{\text{LL}}}{T_{\text{LL}}}\, \text{j}_{\text{LL}},
\end{equation} 
where all observables are evaluated in the Landau--Lifshitz frame. Following App.~\ref{out}, these observables appear as series expansions around their Eckart-frame counterparts, in powers of the 
heat-current norm $\|\text{q}\|$. The latter, displayed below in \eqref{qsq}, is inevitably unbounded for Robinson--Trautman because of the singular future behaviour of $K$. The validity of the hydrodynamic-frame change is therefore limited. This problem has been avoided in our preceding analysis, performed directly and exactly in the original Eckart frame.

Although in Eckart's our choice has been $\mu\equiv\mu_{\text{E}} =0$, this is no longer true in Landau--Lifshitz's (see \eqref{delmT} and \eqref{delmuT2}):\footnote{We use the notation $\text{q}\cdot \tau\cdot \text{q}=\tau_{\mu\nu }\, q^\mu q^\nu$ and similarly for other terms and contractions.}
\begin{equation}
\delta\left(
\frac{\mu}{T} 
\right)= \frac{\text{q}\cdot \tau\cdot \text{q}}{\varrho T \text{q}^2}
- 
 \dfrac{1}{\varrho T
 (p+\varepsilon)}\left(\text{q}^2+\frac{\text{q}\cdot \tau\cdot \tau\cdot \text{q}}{\text{q}^2}-\left(\frac{\text{q}\cdot \tau\cdot \text{q}}{\text{q}^2}\right)^2\right)+
\cdots,
\end{equation} 
where the dots stand for higher-order terms in $\|\text{q}\|$. 
As a consequence, in this frame, the entropy current \eqref{curspLRT} receives two distinct non-vanishing contributions, $\text{S}=\text{S}_{\text{LL}1} +\text{S}_{\text{LL}2}$:
\begin{eqnarray}
&&\text{S}_{\text{LL}1} =s_{\text{LL}} \text{u}_{\text{LL}}=s \text{u} +\frac{s}{p+\varepsilon}\text{q}
-\frac{\mu\varrho\text{q}^2}{T(p+\varepsilon)^2}\text{u}
-s\frac{\tau\cdot \text{q}}{(p+\varepsilon)^2}+\cdots
 ,
\\ 
&&\text{S}_{\text{LL}2}=-\frac{\mu_{\text{LL}}}{T_{\text{LL}}} \text{j}_{\text{LL}}=
\frac{\mu \varrho}{T (p+\varepsilon)}\text{q}+\frac{\mu\varrho\text{q}^2}{T(p+\varepsilon)^2}\text{u}
+s\frac{\tau\cdot \text{q}}{(p+\varepsilon)^2}-\cdots,
\end{eqnarray} 
and we have used the explicit perturbative transformation rules provided in \eqref{delu}--\eqref{delmuT2} (quantities without indices are evaluated in the Eckart frame). These two expressions are general and valid for any fluid. They sum up to $s \text{u} +\nicefrac{\text{q}}{T}$, expression of $\text{S}$ in the Eckart frame.

In the Robinson--Trautman conformal holographic fluid, the heat current $\text{q}$ is given in \eqref{q}:
\begin{equation}
\label{qvec}
\text{q} =-\frac{k^2P^2}{16\pi G}\left(  \partial_{\bar\zeta}K\, \partial_\zeta
+
\partial_\zeta K\,\partial_{\bar\zeta}\right),
\end{equation} 
and its norm squared is
\begin{equation}
\label{qsq}
\text{q}^2 =\frac{k^2P^2}{128\pi^2G^2}\partial_\zeta K\,  \partial_{\bar\zeta}K\, ,
\end{equation} 
while 
\begin{equation}
\label{tauq}
\tau\cdot \text{q}=-\frac{1}{128\pi^2 G^2}\left( \partial_{\bar\zeta}K\, 
\partial_{\zeta} \left(P^2\partial_t\partial_{\zeta}
\ln P
\right)\text{d}\zeta+  \partial_{\zeta}K\, 
\partial_{\bar\zeta} \left(P^2\partial_t\partial_{\bar\zeta}
\ln P
\right)
\text{d}\bar\zeta \right).
\end{equation}
For vanishing chemical potential, $\mu=0$ {(or for vanishing density, $\varrho=0$)}, the above equations read:
\begin{eqnarray}
&&\text{S}_{\text{LL}1} =\text{S}-s\frac{\tau\cdot \text{q}}{(3p)^2}+\cdots,
\label{SLL1ex}
\\ 
&&\text{S}_{\text{LL}2}=s\frac{\tau\cdot \text{q}}{(3p)^2}-\cdots
\label{SLL2ex}
\end{eqnarray} 
with  $p=\nicefrac{\varepsilon}{2}$ given in \eqref{endenRT}, $s$
in
\eqref{entroRT}, $\text{S}$  in \eqref{entrocurRT}
 and  $\tau\cdot\text{q}$ in \eqref{tauq}.
None of the two pieces of the entropy current displayed in the Landau--Lifshitz frame \eqref{SLL1ex} and \eqref{SLL2ex} is divergence-free, but the sum is:
\begin{eqnarray}
\nabla \cdot \text{S}_{\text{LL}1}&=& -\nabla \cdot \text{S}_{\text{LL}2} \nonumber \\
&=& -
\frac{s}{(3p)^2}\nabla\cdot(\tau\cdot \text{q})\nonumber \\
&=&
\frac{P^2}{18k^2(2M)^{\nicefrac{4}{3}}}
\left( \partial_{\zeta}\left(
 \partial_{\zeta}K\, 
\partial_{\bar\zeta} \left(P^2\partial_t\partial_{\bar\zeta}
\ln P
\right)\right)
+ \text{c.c.}\right).
\label{divSLL1}
\end{eqnarray}
In previous analyses of the Robinson--Trautman fluid, $\text{S}_{\text{LL}1}=s_{\text{LL}} \text{u}_{\text{LL}}$ was used alone as an entropy current,
 leading to the conclusion that it is not conserved.\footnote{The expressions for  $\text{S}_{\text{LL}1}$ and
 $\nabla \cdot \text{S}_{\text{LL}1}$ of \cite{Bakas:2014kfa} differ from the ones displayed here, Eqs. \eqref{SLL1ex} and \eqref{divSLL1}, because of 
technical inaccuracies.} This amounts to setting $\mu_{\text{LL}}=0$ in \eqref{curspLRT}, which in turn would
 require $\mu_{\text{E}}\neq 0$. Since in these works no chemical potential was introduced in the original frame
 reached holographically, it seems to us that the choice made subsequently for the entropy current is unjustified. 
Deciding which is the best choice for this current is certainly a long debate
 that we will not pursue here. Our choice is the standard one, originally proposed by Landau and Lifshitz \cite{Landau}. 
{More importantly, it is frame-invariant provided one is careful} in trading the heat current $\text{q}$ for a 
transverse matter current $\text{j}$, when discussing the change of hydrodynamic frame. This is often disregarded 
in the literature.

\section{Conclusions}

We would like now to summarize our analysis, which is twofold.

The first side concerns the general reconstruction of exact bulk Einstein spacetimes, from boundary data obeying appropriate conditions. This reconstruction is a resummation of the hydrodynamic derivative expansion, for which we choose a shearless congruence. Given a boundary metric, such a congruence is basically unique  and has a double virtue: (\romannumeral1)~reducing the number of terms allowed by conformal invariance, hence making the resummation potentially tractable;\footnote{In the presence of shear the plethora of compatible terms makes the exercise difficult.} (\romannumeral2) being promoted into a bulk null, geodesic and shearless congruence, whenever the resummation is successful. This last feature makes the bulk algebraically special by Goldberg--Sachs theorem, and naturally expressed it in Eddington--Finkelstein coordinates. Moreover, it crucially sets a relationship between the boundary energy--momentum tensor and the Cotton tensor, through the structure it imposes on the reference conserved tensors $\text{T}^\pm =\text{T}\pm\frac{i}{8\pi G k^2 }\text{C}$, which is of prime importance. This scheme allows for a direct boundary control of the bulk Petrov type, and recasts the conservation of $\text{T}$ as a bulk integrability equation, interpreted on the boundary as a heat-flow equation. 

The method at hand is general and enables us to reach all known algebraically special Einstein spacetimes (see e.g. \cite{Mukhopadhyay:2013gja, Petropoulos:2015fba} for the Pleba\'nski--Demia\'nski class).   It is fair to quote, though, that the issue of Petrov general spacetimes is still open, together with the r\^ole that a boundary shearless congruence will play in this case, or, stated differently, the possibility of reconstructing such spacetimes with shearless fluid velocities. Leaving aside this question, we have followed the pattern for a general boundary class with a shearless congruence without vorticity and reached the entire Robinson--Trautman family. The Robinson--Trautman equation comes out here holographically as the boundary energy--momentum conservation equation, given the structure the latter acquires from its relationship with the Cotton tensor.

The last property brings us to the second part of the present work, more specifically dedicated to the physics of the holographic fluid. Three main features emerge for it: (\romannumeral1) ~the hydrodynamic frame associated with the congruence at hand is the Eckart frame; 
(\romannumeral2)~in this frame, the energy--momentum tensor receives only third-order derivative corrections; (\romannumeral3) the energy--momentum conservation is non-trivial in the time direction, and appears as the heat equation for the fluid.  These properties can be traced back to our original choice of shearless congruence, and to the consequences it has both for the bulk and for the boundary. They are all expected to be generic for exact and algebraically special Petrov Einstein spaces, and valid beyond the Robinson--Trautman paradigm.

Here, the fluid is at rest on a surface which evolves in time keeping its area constant. The fluid has constant pressure and constant energy density. The transport phenomena occurring can be assimilated with thermal conduction, which drives the system towards global equilibrium by continuously redistributing a conserved total energy on the moving surface, in a fashion reminiscent of Solaris' ocean dynamics \cite{SAT}. This is achieved according to the Calabi flow, here revealed as a genuine heat flow. The interpretation of the Gaussian curvature of the surface as the time-dependent part of a kinematical out-of-equilibrium temperature, and the exact determination of the corresponding geometric heat conductivity are
novelties of our work. They provide a natural thermal-like interpretation to the geometric flow.

The other important aspect unravelled  here concerns the hydrodynamic frame. The holographic fluids dual to exact Einstein (more precisely Einstein--Maxwell in order to produce a boundary current) spacetimes emerge often    
in the Eckart frame. Then, not only is the conserved current perfect, but the corrections to the energy--momentum tensor with respect to the perfect fluid are restricted and canonically related to the third derivatives of the metric and the velocity. This makes the fluid dynamics clear and provides a rich information on series of vanishing transport coefficients. It is unfortunate that in the framework of holography one systematically tries to reach the Landau--Lifshitz frame, irrespective of the  context. This leads sometimes to inconsistencies, as we pointed out e.g. regarding the entropy current.

The present analysis of the Robinson--Trautman boundary fluid, and other studies of exact-Einstein-space holography, suggest that the underlying fluid dynamics is quite peculiar.  The system is time-dependent and evolves generically towards equilibrium by thermal conduction. This process is of geometric origin though, as it is driven by the evolution of the surface itself, and is associated to a very specific correction with respect to perfect fluidity. Furthermore energy and area are conserved, and the standard entropy current has no divergence. Entropy is thus conserved as a fine tuning inside the out-of equilibrium process at hand. There is nothing to be worried about this state of affairs, except that  one might legitimately question the practical usefulness of these holographic systems and the interest in elaborating further on their transport properties. In contrast, the investigation of this distinctive conformal fluid dynamics, might shed light on black-hole out-of-equilibrium thermodynamics, which is still in a quite primitive state.

\section*{Acknowledgements}

This work is partly based on a talk delivered during Corfu summer institute 2016 
\textsl{School and workshops on elementary particle physics and gravity}. It relies on works which are 
published or will appear in collaboration with J. Gath, C. Marteau and A. Mukhopadhyay.
We would like to thank them, together with M. Appels, R. Balian, G. Barnich, Q. Bonnefoy, S. Detournay, P. Kovtun, R. Leigh, O. Miskovic and R. Olea for useful correspondence and valuable discussions. 
Marios Petropoulos would like to thank the Universit\`a di Torino, the Universidad Andr\'es Bello de Santiago de Chile and the Pontificia Universidad Cat\'olica de Valparaiso, where parts of the present work were also presented. We thank each others home institutions for hospitality and financial support.

\appendix 
\section{On vector-field congruences}  \label{cong}

Consider a $D$-dimensional Lorentzian metric $g_{\mu\nu}$ and an arbitrary time-like vector
 field $\text{u}=u^\mu \partial_\mu$, normalized as  $u_\mu u^\mu=-1$,  later identified 
with the fluid velocity. Its integral curves define a congruence which is characterized by
 its acceleration, shear, expansion and vorticity:
\begin{equation}
\label{def1}
\nabla_{\mu } u_\nu =-u_\mu  a_\nu  +\frac{1}{D-1}\Theta\, h_{\mu \nu}+\sigma_{\mu \nu} +\omega_{\mu \nu}
\end{equation}
with\footnote{Our conventions for symmetrization and antisymmetrization are: 
$$A_{(\mu \nu)}=\frac{1}{2}\left(A_{\mu \nu}+A_{\nu \mu}\right)\,,\quad 
A_{[\mu \nu] }=\frac{1}{2}\left(A_{\mu \nu}-A_{\nu \mu}\right).$$}
\begin{eqnarray}
a_\mu &=&u^\nu \nabla_\nu u_\mu , \quad
\Theta=\nabla_\mu  u^\mu , \label{def21}\\
\sigma_{\mu \nu}&=&\frac{1}{2} h_\mu ^{\hphantom{\mu } \rho}  h_\nu ^{\hphantom{\nu }\sigma}\left(
\nabla_\rho  u_\sigma+\nabla_\sigma u_\rho
\right)-\frac{1}{D-1} h_{\mu \nu}h^ {\rho \sigma} \nabla_\rho u_\sigma \label{def22}\\
&=& \nabla_{(\mu } u_{\nu )} + a_{(\mu } u_{\nu )} -\frac{1}{D-1} \Theta\,h_{\mu \nu}  ,
\label{def23}\\
\omega_{\mu \nu}&=&\frac{1}{2} h_\mu ^{\hphantom{\mu } \rho}  h_\nu ^{\hphantom{\nu }\sigma}\left(
\nabla_\rho u_\sigma-\nabla_\sigma u_\rho
\right)= \nabla_{[\mu } u_{\nu ]} + u_{[\mu } a_{\nu] }.
\label{def24}
\end{eqnarray}
These tensors satisfy several simple identities:
\begin{equation}
u^\mu  a_\mu =0, \quad u^\mu  \sigma_{\mu \nu}=0,\quad u^\mu  \omega_{\mu \nu}=0, 
\quad u^\mu  \nabla_\nu  u_\mu =0, \quad h^\rho_{\hphantom {\rho] }\mu} \nabla_\nu  u_\rho =\nabla_\nu  u_\mu ,
\end{equation}
and we have introduced the longitudinal and transverse projectors:
\begin{equation}
\label{proj}
U^\mu _{\hphantom{\mu }\nu} = - u^\mu  u_\nu , \quad h^\mu _{\hphantom{\mu }\nu} =  u^\mu  u_\nu  + \delta^\mu _{\nu},
\end{equation}
where $h_{\mu \nu}$ is also the induced metric on the local plane orthogonal  to $\text{u}$. 
The projectors satisfy the usual identities:
\begin{equation}
U^\mu _{\hphantom{\mu } \rho}  U^\rho_{\hphantom {\rho] }\nu} = 
U^\mu _{\hphantom{\mu }\nu},\quad U^\mu _{\hphantom{\mu } \rho}  h^\rho_{\hphantom {\rho] }\nu}  =   0 ,
 \quad h^\mu _{\hphantom{\mu } \rho}  h^\rho_{\hphantom {\rho] }\nu}  =   h^\mu _{\hphantom{\mu }\nu} ,\quad 
U^\mu _{\hphantom{\mu }\mu}=1, \quad h^\mu _{\hphantom{\mu }\mu}=D-1.
\end{equation}

It is customary to define the orthogonally projected covariant derivative acting on any tensor as 
\begin{equation}
\label{nablaproj}
D_\gamma T_{\alpha_1\ldots \alpha_p}^{\hphantom{\alpha_1\ldots \alpha_p}\beta_1\ldots \beta_q} =
h_\gamma^{\hphantom{\gamma}\lambda} h_{\alpha_1}^{\hphantom{\alpha_i}\mu_1}\dots  
h_{\alpha_p}^{\hphantom{\alpha_i}\mu_p}h^{\hphantom{\nu_i}\beta_1}_{\nu_1}\dots  h^{\hphantom{\nu_i}\beta_q}_{\nu_q}
\nabla_\lambda T_{\mu_1\ldots \mu_p}^{\hphantom{\mu_1\ldots \mu_p}\nu_1\ldots \nu_q}.
\end{equation}

Any tensor can be decomposed in longitudinal, transverse and mixed components. Consider for concreteness 
the energy--momentum tensor, which is rank-two and symmetric with components $T_{\mu \nu}$:
\begin{equation}\label{Tdecg} 
T_{\mu \nu}=\pmb{\varepsilon} u_\mu  u_\nu +\pmb{p}  h_{\mu\nu}+   \tau_{\mu \nu}+u_\mu  q_\nu  + u_\nu  q_\mu .
\end{equation}
The non-longitudinal part is 
\begin{equation}\label{Sdec}
\pmb{p}  h_{\mu\nu}+   \tau_{\mu \nu}+u_\mu  q_\nu  + u_\nu  q_\mu .
\end{equation}
We have defined
\begin{equation}
\pmb{\varepsilon} = u^\mu  u^\nu  T_{\mu \nu},\quad
     \tau_{\mu \nu}=  h_\mu ^{\hphantom{\mu } \rho}  h_\nu ^{\hphantom{\nu}\sigma} T_ {\rho \sigma}-\pmb{p}  h_{\mu\nu},\quad
q_\mu = - h_\mu ^{\hphantom{\mu }\nu}T_{\nu  \sigma} u^\sigma
\end{equation}
such that 
\begin{equation}\label{trans}
h_\mu ^{\hphantom{\mu }\nu}  q_\nu =q_\mu , \quad h_\mu ^{\hphantom{\mu }\rho}    \tau_{\rho\nu}= \tau_{\mu\nu}, \quad
u^\mu  q_\mu =0, \quad u^\mu    \tau_{\mu \nu}=0, \quad
u^\mu  T_{\mu \nu}=- q_\nu -\pmb{\varepsilon} u_\nu .
\end{equation}
The purely transverse piece $  \pmb{p}h_{\mu\nu}+   \tau_{\mu \nu}$ is  the stress tensor, while $q^\mu$ is the heat current. 

Similarly, any current with components $J^\mu$ can be decomposed in longitudinal and transverse parts:
\begin{equation}
\label{curdec}
J^\mu=\pmb{\varrho} u^\mu +j^\mu 
\end{equation}
with
\begin{equation}\label{transcurr}
h_\mu ^{\hphantom{\mu }\nu}  j_\nu =j_\mu , \quad u^\mu  j_\mu =0, \quad 
\pmb{\varrho} = - u^\mu J_\mu.
\end{equation}

Assuming the energy--momentum tensor $T_{\mu \nu}$ being conserved:
\begin{equation}\label{Tconser}
\nabla_\mu T^{\mu\nu}=0,
\end{equation}
we can carry on and describe the dynamics for the heat current, using \eqref{trans}, together with 
\eqref{def1} and \eqref{Tdecg}. We obtain, for its divergence:\footnote{Notice that $\text{q}$ being transverse,
$D_\mu q^\mu=\nabla_\mu q^\mu-a_\mu q^\mu$.}
\begin{equation}\label{divq}
\nabla_\mu q^\mu= -\text{u}(\pmb{\varepsilon})-
\left(\pmb{p}+\pmb{\varepsilon}+\frac{g_{\mu\nu} \tau^{\mu\nu}}{D-1}\right) \Theta- a_\mu q^\mu-\sigma_{\mu\nu}\tau^{\mu\nu},
\end{equation}
where
$\text{u}(f)=u^\mu \nabla_\mu(f) =u^\mu \partial_\mu(f)$.

The current $\text{J}$ is also supposed to to obey 
\begin{equation}\label{Jconser}
\nabla_\mu J^{\mu}=0,
\end{equation}
from which we extract the dynamics of its transverse component $\text{j}$ using \eqref{def21}:
\begin{equation}\label{divu}
\nabla_\mu j^\mu= -\text{u}(\pmb{\varrho})-
\pmb{\varrho}\,\Theta.
\end{equation}

\section{Hydrodynamics and out-of-equilibrium states}  \label{out}

\subsection*{Hydrodynamic functions  and hydrodynamic frames}

We recall here some basic facts regarding fluid dynamics (see \cite{Kovtun:2012rj,Romatschke:2009im} as well as the pillar of hydrodynamics manuals \cite{Landau} -- we also recommend \cite{RZ}). Hydrodynamics is by essence out-of-equilibrium. Every concept should therefore be 
considered with care, as no universal methods exist, which would embrace all facets of these phenomena, 
especially in the relativistic regime for non-ideal fluids. 

Fluids are described in terms of their energy--momentum tensor and one (or more) current(s), 
all conserved in the absence external forces. The  dynamical quantities are thus (see \eqref{Tdecg} 
and \eqref{curdec}) $\pmb{\varepsilon}(x)$, $\pmb{p}(x)$, $\pmb{\varrho}(x)$, $q^\mu(x)$, $\tau^{\mu\nu}(x)$ 
and $j^\mu(x)$, assumed to be functionals of some fundamental quantities, equal in number to the available 
equations \eqref{Tconser} and \eqref{Jconser}: $u^\mu(x)$, $\pmb{T}(x)$ and $\pmb{\mu}(x)$. This functional 
dependence is captured by the constitutive equations, expressed usually as a derivative 
expansion. As a matter of principle, these hydrodynamic functionals obey microscopic equations
like Boltzmann's equation, but it is in practice difficult to extract information directly from there. 
The derivative expansion is the alternative, perturbative phenomenological approach.

At strict equilibrium and for an ideal fluid, $\text{u}$ is aligned with a time-like Killing vector, \emph{i.e.} 
the fluid is at rest, and $\pmb{T}$ and $\pmb{\mu}$ are constants. So are $\pmb{\varepsilon}$, $\pmb{p}$ 
and $\pmb{\varrho}$. All these quantities are then defined  within equilibrium thermodynamics as the 
temperature $T$, chemical potential $\mu$, energy density $\varepsilon$, pressure $p$ and matter
 (or better, N\oe ther-charge) density $\varrho$. The constitutive relations are the equation of state $p=p(T,\mu)$
and the usual Gibbs--Duhem relation for the grand potential $-p=\varepsilon-Ts-\mu\varrho$ with  
$\varrho=\left(\nicefrac{\partial p}{\partial \mu}\right)_T$ and  $s=\left(\nicefrac{\partial p}{\partial T}\right)_\mu$.

Once the fluid is set to motion, the equilibrium is abandoned and assumed to be achieved locally, 
for hydrodynamics to make sense. Thermodynamic functions become local (and supposed to be slowly varying)
 but even within this basic assumption, for non-perfect fluids, neither  $\pmb{\varepsilon}(x)$, 
$\pmb{p}(x)$ and $\pmb{\varrho}(x)$ appearing in the fluid equations, nor  $\pmb{T}(x)$ and $\pmb{\mu}(x)$ 
entering the constitutive relations need \emph{a priori} to be identified with the corresponding local-equilibrium
 thermodynamic quantities. Even the velocity congruence $\text{u}(x)$ has no first-principle definition
 in relativistic hydrodynamics. 
One has in particular the freedom to redefine 
\begin{equation}
\pmb{T}(x)\to\pmb{T}^{\prime}(x),\quad \pmb{\mu}(x)\to\pmb{\mu}^{\prime}(x), \quad \text{u}(x)\to\text{u}^{\prime}(x),
\end{equation} 
provided we modify accordingly $\pmb{\varepsilon}(x)$, $\pmb{p}(x)$, $\pmb{\varrho}(x)$, 
$q^\mu(x)$, $\tau^{\mu\nu}(x)$ and $j^\mu(x)$. 

The above freedom can be used to fix some of the hydrodynamic functions. This is how the
 concept of hydrodynamic frame emerges. The Eckart frame (also called particle frame,  \cite{preEckart, Eckart}) is reached by requiring the matter current 
$\text{J}$ be perfect \emph{i.e.} $\text{j}=0$, while in the Landau--Lifshitz frame the 
heat current $\text{q}$ is set to zero \cite{Landau}. In every frame, the remaining non-vanishing 
hydrodynamic functionals are set as derivative expansions with respect to 
$\pmb{T}(x)$, $\pmb{\mu}(x)$ and $u^\mu(x)$. The coefficients are phenomenological data, 
which can in principle be determined from the microscopic theory. The consequence of 
changing frame is to reshuffle the various coefficients (sometimes trading one for an 
infinite number of others), which ultimately carry the relevant information about the fluid,
 irrespective of the frame.
 
It is worth noting at this stage that the definition of the Eckart frame and, by the logic of frame transformation, 
the corresponding definition of the Landau--Lifshitz counterpart, refer \emph{explicitly} to the conserved matter current
 $\text{J}$. 
{The heat current $\text{q}$, as part of the conserved energy--momentum 
tensor $\text{T}$, and the non-perfect contribution $\text{j}$ to the conserved matter current $\text{J}$}
are interchanged in the course of the transformation. Regularity (or invertibility) of the latter makes it dangerous
 to set \emph{a priori} both these vectors to zero, irrespective of the fact that ultimately the matter 
density $\pmb \varrho$ or the chemical potential $\pmb \mu$  may vanish.

The choice of frame is important for several reasons. At the first place,
 because of the nature of derivative expansions: these are often asymptotic series and only the 
first terms can  be trusted. Hence, depending on the regime, some frames may not provide accurate 
results. Secondly, the precise physical context can play a r\^ole. For instance, when dealing with 
fluids in a quasi-Newtonian regime, the Eckart frame is superior as it is the one in which one recovers 
classical Euler's equations for non-relativistic fluids. Following the classical irreversible thermodynamics theory in 
Eckart frame,\footnote{See \cite{RZ} for a comprehensive 
review about classical irreversible thermodynamics (CIT) and the Eckart frame.} we find at 
first order -- dropping the index ``E'':
\begin{eqnarray}\label{e1}
&&\pmb{\varepsilon}_{(1)}=\varepsilon,\quad\pmb{p}_{(1)}=p,\quad \pmb{\varrho}_{(1)}=\varrho,\\
\label{s1}&&\tau_{(1)}^{\mu \nu}=-2\eta \sigma^{\mu \nu}-\zeta h^{\mu\nu}\Theta,\\
\label{q1} &&q_{(1)}^{\mu}= -\kappa h^{\mu\nu}\left(\partial_\nu T+T\, a_\nu \right).
\end{eqnarray}
In $D=3$ spacetime dimensions there is also a 
term $ -\zeta_{\mathrm{H}}\eta_{\vphantom{\lambda}}^{\rho\lambda(\mu}u_\rho\sigma_\lambda^{\hphantom{\lambda}\nu)}$ 
in $\tau_{(1)}^{\mu \nu}$ with $\zeta_{\mathrm{H}}$ the Hall viscosity.

Formally, the choice of frame (Eckart, Landau--Lifshitz, \dots) does not exhaust all freedom and it is 
always implicitly assumed that, owing to this residual latitude, $\pmb{\varepsilon}(x)$, $\pmb{p}(x)$  
and $\pmb{\varrho}(x)$ are identified with the local-equilibrium thermodynamic energy density  $\varepsilon(x)$, 
pressure $p(x)$ and charge density $\varrho(x)$, \emph{i.e.} not only at the first order as Eqs. \eqref{e1} 
may suggest.  Nothing guarantees, however, that the kinematic out-of-equilibrium temperature $\pmb{T}(x)$
 and chemical potential $\pmb{\mu}(x)$ could be identified with the equilibrium data $T(x)$ and  $\mu(x)$, 
even at lowest order -- \emph{a fortiori} when higher (and possibly all) orders in the derivative expansion 
are concerned. The literature is very poor on this issue, probably because we are here reaching the limits 
of the hydrodynamic approach. Answering this question would require to enter the realm of non-equilibrium many-body systems.

\subsection*{Conformal fluids}
 
The case of conformal fluids deserves some further comments. From microscopic first principles, 
the energy--momentum tensor is traceless and this should hold even in the limit of extinct interactions. 
In other words, from Eq. \eqref{Tdecg} and following the above identification of kinematical energy and 
pressure  $\pmb{\varepsilon}$, $\pmb{p}$ with thermodynamic ones $\varepsilon$, $p$, one obtains:
\begin{equation}
 \varepsilon(x) = \left(D-1\right)p(x),\quad g_{\mu\nu} \tau^{\mu\nu}=0.
 \label{conf}
\end{equation} 
Equilibrium thermodynamics for conformal fluids then sets the equilibrium temperature $T(x)$ following Stefan's law,  modified in the presence of a chemical potential to comply with the Gibbs--Duhem equation \eqref{thermorel}:
\begin{equation}
\label{stefan}
p = T^D f\left(\frac{\mu}{T}\right).
\end{equation} 
Here $f\left(\nicefrac{\mu}{T}\right)$ encodes the equation of state for the conformal fluid. It is determined by its microscopic properties, and satisfies 
\begin{equation}
\label{stefanorig}
f(0) =\frac{\sigma}{D-1},
\end{equation} 
where $\sigma$ is a Stefan--Boltzmann-like constant in $D$ dimensions. The matter density and entropy therefore read:
\begin{eqnarray}
\varrho&=&\left(\frac{\partial p}{\partial \mu}\right)_T=T^{D-1}f'\left(\frac{\mu}{T}\right),
\\ 
s&=&\left(\frac{\partial p}{\partial T}\right)_\mu=\frac{1}{T}(Dp-\mu\varrho).
\end{eqnarray} 
Vanishing density requires thus $f=\nicefrac{\sigma}{D-1}$ constant, and we recover Stefan's law in this case too. As already emphasized, the thermodynamic temperature and chemical potential may not be meaningful in a plain non-equilibrium regime.

\subsection*{Entropy current}

The next object we would like to discuss is the entropy current. The canonical  expression for it  is
 \cite{Kovtun:2012rj,RZ,Landau, Israel}
\begin{equation}
\label{curs}
S^\mu=\frac{1}{T}\left(p u^\mu-T^{\mu\nu}u_\nu-\mu J^\mu
\right).
\end{equation} 
Using the decompositions \eqref{Tdecg} and \eqref{curdec}, the identifications of the kinematical 
 $\pmb{\varepsilon}(x)$, $\pmb{p}(x)$  and $\pmb{\varrho}(x)$ with the thermodynamic ones, as well 
as the already quoted equilibrium thermodynamic relation 
\begin{equation}
\label{thermorel}
Ts=p+\varepsilon-\mu\varrho,
\end{equation}  
one finds:
\begin{equation}
\label{cursp}
S^\mu=s u^\mu+\frac{1}{T}q^\mu-\frac{\mu}{T} j^\mu.
\end{equation} 
This current allows writing the thermodynamic entropy as:
\begin{equation}
\label{entrocurr}
s=-S^\mu u_\mu.
\end{equation} 
We should stress that the entropy current has raised many questions and its canonical form  \eqref{curs} may not be appropriate to all physical situations. It is  
based on local-equilibrium thermodynamic functions, $s(x)$, $T(x)$ and $\mu(x)$, and depending on the set-up, these may be far from the kinematical 
$\pmb{T}(x)$ and $\pmb{\mu}(x)$, which lack first-principle microscopic definition anyway.

It can be shown that the entropy current is frame-independent \cite{Kovtun:2012rj}. This holds in 
particular for Eckart and Landau--Lifshitz frames:
\begin{equation}
\label{curspELL}
S^\mu_{\text{E}}=S^\mu_{\text{LL}}.
\end{equation} 
The formal expression of the current changes though, from one frame to another.  In the Eckart 
frame, \eqref{curs} becomes
\begin{equation}
\label{curspE}
S^\mu_{\text{E}}=s u^\mu+\frac{1}{T}q^\mu,
\end{equation} 
and using Eq. \eqref{divq}
\begin{eqnarray}
\nonumber
\nabla_\mu S^\mu_{\text{E}}
&=&- \frac{\mu \varrho}{T}\Theta-\text{u}\left(\frac{\mu \varrho}{T}\right) +\text{u}\left(\frac{p}{T}\right) 
+\varepsilon\text{u}\left(\frac{1}{T}\right) +\text{q}\left(\frac{1}{T}\right)
\nonumber \\
&&-\frac{1}{T}\left(
\frac{g_{\mu\nu} \tau^{\mu\nu}}{D-1}\Theta + a_\mu q^\mu+\sigma_{\mu\nu}\tau^{\mu\nu}
\right)
.\label{curspEN}
\end{eqnarray} 
Similarly, we find in the Landau--Lifshitz frame 
\begin{equation}
\label{curspL}
S^\mu_{\text{LL}}=s u^\mu-\frac{\mu}{T} j^\mu,
\end{equation} 
which is precisely the current originally proposed by Landau and Lifshitz in \cite{Landau}. Thanks to the usual tools (\eqref{curdec},  \eqref{Jconser} and \eqref{divu}), the divergence turns out to be
\begin{equation}
\nabla_\mu S^\mu_{\text{LL}}=
\text{u}\left(\frac{p+\varepsilon}{T}\right) 
+\frac{p+\varepsilon}{T}\Theta
-\varrho\text{u}\left(\frac{\mu }{T}\right) 
-\text{j}\left(\frac{\mu }{T}\right)
.\label{curspLN}
\end{equation} 
In order to avoid cluttering indices, it is understood that whatever quantity appears in the right-hand side of Eqs. \eqref{curspE}--\eqref{curspLN} is 
determined in the hydrodynamic frame declared in the  left-hand side (and similarly for  \eqref{curspERTcharged}--\eqref{curspconfL} below).

Positivity of $\nabla_\mu S^\mu$ sets bounds on the transport 
coefficients that appear in the derivative expansion. Notice \emph{en passant} that this divergence is Weyl-covariant  
as it matches the Weyl-divergence of the entropy current.\footnote{Indeed, we would write
$\mathscr{D}_\mu S^\mu=\nabla_\mu S^\mu+(w_{\text{S}}-D)A_\mu S^\mu
$, but the conformal weight $w_{\text{S}}$ of the entropy current equals $D$. }

For a conformal fluid, the entropy current \eqref{cursp} reads:
\begin{equation}
\label{curspERTcharged}
S^\mu_{\text{E}}=\frac{1}{T}\big((Dp-\mu\varrho) u^\mu+q^\mu\big),\quad \text{or} \quad
S^\mu_{\text{LL}}=\frac{1}{T}\big((Dp-\mu\varrho) u^\mu-\mu j^\mu\big),
\end{equation} 
while its divergence \eqref{curspEN} or \eqref{curspLN} is now
\begin{equation}
\nabla_\mu S^\mu_{\text{E}}=
- \frac{\mu \varrho}{T}\Theta-\text{u}\left(\frac{\mu \varrho}{T}\right)+
D p\,\text{u}\left(\frac{1}{T}\right) +\text{q}\left(\frac{1}{T}\right)
-\frac{1}{T}\left(a_\mu q^\mu+\sigma_{\mu\nu}\tau^{\mu\nu}
-\text{u}(p) 
\right),
\label{curspconf}
\end{equation} 
or
\begin{equation}
\nabla_\mu S^\mu_{\text{LL}}=
D\,\text{u}\left(\frac{p}{T}\right) 
+D\frac{p}{T}\Theta
-\varrho\text{u}\left(\frac{\mu }{T}\right) 
-\text{j}\left(\frac{\mu }{T}\right).
\label{curspconfL}
\end{equation} 
The various kinematical and thermodynamic quantities appearing in the equations, are determined in the corresponding frame; {they are different for Eckart and Landau--Lifshitz, contrary to the entropy current and its divergence}.

\subsection*{Eckart-to-Landau--Lifshitz transformation}

We would like to conclude this appendix with some explicit  transformation rules. Writing $\mathscr{Q}_{\text{LL}}=\mathscr{Q}_{\text{E}} +\delta \mathscr{Q}$ for any kinematical or thermodynamic quantity $\mathscr{Q}$, the displacements can be computed linearly, quadratically, and so on, based on the fundamental rule that the energy--momentum tensor $\text{T}$ and the matter current $\text{J}$ are frame-invariant. In order to avoid any confusion, we restore the index ``E'' for the Eckart frame, and provide the results with minimal details.

The variation in the velocity field is determined in terms of the heat current, non-zero in Eckart frame, vanishing in Landau--Lifshitz frame, by solving perturbatively the eigenvalue problem:
\begin{equation}
\label{delu}
\delta \text{u}^{(1)}=\frac{\text{q}}{p_\text{E} + \varepsilon_\text{E} } .
\end{equation}
All other transformation rules are determined from the latter, using the quoted invariances and Gibbs--Duhem equation.\footnote{The kinematical  $\pmb{\varepsilon}_{\text{LL}}(x)$, $\pmb{p}_{\text{LL}}(x)$  
and $\pmb{\varrho}_{\text{LL}}(x)$ are still identified with the local-equilibrium thermodynamic energy density  $\varepsilon_{\text{LL}}(x)$, 
pressure $p_{\text{LL}}(x)$ and charge density $\varrho_{\text{LL}}(x)$.}  The non-perfect matter-current component $\text{j}$ is vanishing in Eckart and non-zero in Landau--Lifshitz, where its first-order value is 
\begin{equation}
\label{delj1}
\delta \text{j}^{(1)}=-\frac{\varrho_\text{E}}{p_\text{E} + \varepsilon_\text{E} } \text{q}\,,
\end{equation}
while 
\begin{equation}
\label{dele}
\delta \varepsilon^{(1)}=\delta \varrho^{(1)}=\delta s^{(1)}=\delta p^{(1)}=0\,.
\end{equation}
Similarly, we find
\begin{equation}
\label{delmT}
\delta\left(\frac{\mu}{T}\right)^{(1)}=\frac{q\cdot \tau_{\text{E}}\cdot q}{\varrho_\text{E}  T_\text{E}  \text{q}^2},
\end{equation}
and using $\delta p = \varrho \delta \mu +s \delta T$ we can read off 
$\delta T^{(1)}$ and $\delta \mu^{(1)}
$. 

It should be noticed that the stress tensor $\tau_\text{E}$ is a correction with respect to the perfect fluid, of similar order than the heat current $\text{q}$. The first correction it receives is therefore of second order:
\begin{equation}
\label{deltau}
\delta \tau^{(2)\mu\nu }=\frac{q\cdot \tau_{\text{E}}\cdot q}{\left(p_\text{E}+  \varepsilon_\text{E}\right)  \text{q}^2}\left(q^\mu u^\nu+q^\nu u^\mu\right)+ \frac{\text{tr}\, \delta \tau^{(2)}}{D-1}h^{\mu\nu}.
\end{equation}
In this expression, the trace of the correction, $\text{tr}\, \delta \tau^{(2)}=g_{\mu\nu}\, \delta \tau^{(2)\mu\nu}$, is left undetermined. This trace also appears in the second-order correction of the pressure,
\begin{equation}
\label{delp2}
\delta p^{(2)} = \dfrac{\delta \varepsilon^{(2)}}{D-1}
-\dfrac{\text{tr}\, \delta \tau^{(2)}}
{D-1}, \quad \delta \varepsilon^{(2)} = -\dfrac{\text{q}^2}{p_\text{E}+\varepsilon_\text{E}},
\end{equation}
so that a freedom remains to reabsorb it or not in the latter (see discussion in \cite{Kovtun:2012rj}). 
The other second-order corrections from Eckart to Landau--Lifshitz frame read:
\begin{eqnarray}
\label{delu2}
&\delta \text{u}^{(2)} = \dfrac{1}{2\left(p_\text{E}+\varepsilon_\text{E}\right)^2}\left(\text{q}^2\text{u}_{\text{E}} - 2\tau_{\text{E}}\cdot \text{q}_{\text{E}}\right),
&\\
\label{delj2}
&\delta \text{j}^{(2)} =- \dfrac{\varrho_{\text{E}}}{\left(p_\text{E}+\varepsilon_\text{E}\right)^2}\left(\text{q}^2\text{u}_{\text{E}} 
- \tau_{\text{E}}\cdot \text{q}_{\text{E}}\right),
&\\
\label{dels2}
&\delta s^{(2)} =  \dfrac{\text{q}^2\text{s}_{\text{E}}}{2\left(p_\text{E}+\varepsilon_\text{E}\right)^2}-  \dfrac{\text{q}^2}{T_{\text{E}}\left(p_\text{E}+\varepsilon_\text{E}\right)},&
\\&
\delta \varrho^{(2)} =  \dfrac{\text{q}^2\varrho_{\text{E}}}{2\left(p_\text{E}+\varepsilon_\text{E}\right)^2},
&\\
 \label{delmuT2}
&\delta{\left(\dfrac{\mu}{T}\right)}^{(2)} = - 
 \dfrac{1}{\varrho_{\text{E}} T_{\text{E}} 
 \left(p_\text{E}+\varepsilon_\text{E}\right)}
 \left(\text{q}^2+\dfrac{\text{q}\cdot \tau_{\text{E}}\cdot \tau_{\text{E}}\cdot \text{q}}{\text{q}^2}-\left(\dfrac{\text{q}\cdot \tau_{\text{E}}\cdot \text{q}}{\text{q}^2}\right)^2\right) .&
\end{eqnarray}
Finding the latter requires to analyse the eigenvalue problem of the energy--momentum tensor at third order.  We can further combine \eqref{delp2} with \eqref{delmuT2} and $\delta p = \varrho \delta \mu +s \delta T$,  and extract $\delta T^{(2)}$ and $\delta \mu^{(2)}$.

We can proceed similarly and obtain the above quantities at next order, or even further. Their expressions follow the pattern already visible in the first and second orders.
It is readily seen that the expansions of all Landau--Lifshitz observables around their Eckart values are controlled by the parameter $\nicefrac{\|\text{q}\|}{p_\text{E} + \varepsilon_\text{E}}$, \emph{i.e.} basically the norm of the heat current. The magnitude of this quantity sets validity bounds on the frame transformation at hand. For a more general discussion on related issues, see the already quoted Refs. \cite{Kovtun:2012rj, RZ}.


\end{document}